\documentclass[aip,preprint]{revtex4-1}
\usepackage[english]{babel}
 
\usepackage{comment} 
\usepackage{bm}%
\usepackage{xspace}

\usepackage{graphicx}
\usepackage{subfig}

\usepackage[version=3]{mhchem}
\usepackage{amsmath}

\usepackage[normalem]{ulem}

\newcommand{\cunh}[0]{[\text{Cu(NH}_3)_4]^{2+}}
\newcommand{\cuwater}[0]{[\text{Cu(H}_2\text{O})_4]^{2+}}
\newcommand{\cucl}[0]{[\text{CuCl}_4]^{2-}}

\newcommand{\ket}[1]{{\ensuremath{|#1\rangle}\xspace}}
\newcommand{\bra}[1]{{\ensuremath{\langle #1|}\xspace}}

\newcommand{\lmct}[0]{{\ensuremath{\ket{\rm LMCT}\xspace \,}}}

\newcommand{\elemm}[3]{{\ensuremath{\bra{#1}{#2}\ket{#3}}\xspace}}
\newcommand{\ovrlp}[2]{{\ensuremath{\langle #1|#2\rangle}\xspace}}

\title{Interplay between electronic correlation and metal-ligand delocalization in the spectroscopy of transition metal compounds: case study on a series of planar Cu$^{2+}$ complexes.} 

\begin{document}
\author{Emmanuel Giner}%
\email{emmanuel.giner@lct.jussieu.fr}
\affiliation{Laboratoire de Chimie théorique, Sorbonne Université, UMR 7616,
4 place Jussieu, 75252 Paris, France}
\author{David Tew}
\affiliation{Max Planck Institute for Solid State Research, Heisenbergstra$\beta$e 1, 70569 Stuttgart}
\author{Yann Garniron}
\affiliation{Laboratoire de Chimie et Physique Quantique, UMR 5626, Université Paul Sabatier, 118 route de Narbonne 31062 Toulouse, France}
\author{Ali Alavi}
\affiliation{Max Planck Institute for Solid State Research, Heisenbergstra$\beta$e 1, 70569 Stuttgart}

\begin{abstract}
We present a comprehensive theoretical study of the physical phenomena that determine the relative energies of the three of the lowest
electronic states of each of the square-planar copper complexes $\cucl$, $\cunh$ and $\cuwater$, and present a detailed analysis of 
the extent to which truncated configuration interaction (CI) and coupled cluster (CC) theories succeed in predicing the excitation energies.
We find that ligand-metal charge transfer (CT) single excitations play a crucial role in the correct determination of the properties of these systems, 
even though the CT processes first occur at fourth order in perturbation theory, and  
propose a suitable choice of minimal active space for describing these systems with multi-reference theories.
CCSD energy differences agree very well with near full CI values even though the $T_1$ diagnostics are large, which casts doubt on the usefulness
of singles-amplitude based multi-reference diagnostics.
CISD severely underestimates the excitation energies and the failure is a direct consequence of the size-inconsisency errors in CISD.
Finally, we present reference values for the energy differences computed using explicitly correlated CCSD(T) and BCCD(T) theory.
\end{abstract}

\maketitle

\section{Introduction}

Open-shell transition metal complexes, which are ubiquitous in biological and industrial chemistry, represent one of the
main challenges for present-day quantum chemistry, where theory seeks to provide prediction and interpretation of 
key properties such as electronic transition energies, spin-density maps and magnetic anisotropy.
Complexes containing Cu$^{2+}$ have been studied extensively,
both using Density Functional and wavefunction theories,\cite{PhysRevB_jansen,vanOosten1996_noci,carmen_jp_2000_tab,caspt2_degraaf,malrieu_magnet_1,carmen_jp_2002_deloc_nos,Broer_noci2002,solomon_cucl4,neese_CuNH3_SOS,ramirez_daudey_cucl2,atanasov_cucl4,malrieu_magnet_3,neese_epr_g_tensor_dft,g_tensor_pierloot_mscaspt2,dmrg_sd,dft_sd_boguslawski,cucl2_1,cucl2_cele} and have been found to pose a tough test for electronic structure methods.
Popular functionals such as B3LYP and BP86 systematically underestimate the spin-density at the Cu atom, provide
poor $d-d$ and ligand-to-metal exctiation energies\cite{ramirez_daudey_cucl2,atanasov_cucl4,solomon_cucl4} and
misleading predictions of magnetic anisotropy tensors.\cite{neese_epr_g_tensor_dft,neese_CuNH3_SOS,Remenyi_cucl4,solomon_cucl4}
Although it is possible to design tailored functionals for these systems, with higher percentages of Hartree--Fock exchange, this pragmatic
approach has limited transferability and limited predictive power.

On the other hand, studies using wavefunction methods have also only been partially successful. 
Transtition metal complexes are considered to be strongly correlated systems and 
Complete Active Space Self Consistent Field (CASSCF) theory is usually applied, with multirefernce perturbation or truncated CI corrections for 
dynamic correlation. The computed energies are found to be highly sensitive to the choice of active space and 
the level of coupling between the treatment of static and dynamic correlation, but the number of orbitals involved in the coordination
at the transition metal centre prohibits brute force convergence with respect to the size of the active space.
Although the relatively high density of low lying electronic states and the large values of $T_1$ diagnostics 
observed for transition metal complexes discourages the use of single reference methods, the accuracy of
single reference coupled-cluster methods for these sytems remains an open question.
This paper reports the results of a series of careful benchmark calculations and detailed theoretical analysis, performed on 
three Cu$^{2+}$ complexes $\cucl$, $\cunh$ and $\cuwater$. We address the question of what characteristics
an electronic structure method should have in order to correctly describe the lower lying electronic states and the
spin-densities at the Cu atom and analyse the sucesses and failures of commonly applied single-reference and 
multi-reference wavefunction methods.

\begin{figure}[h]
\subfloat[]{\includegraphics[scale=0.40]{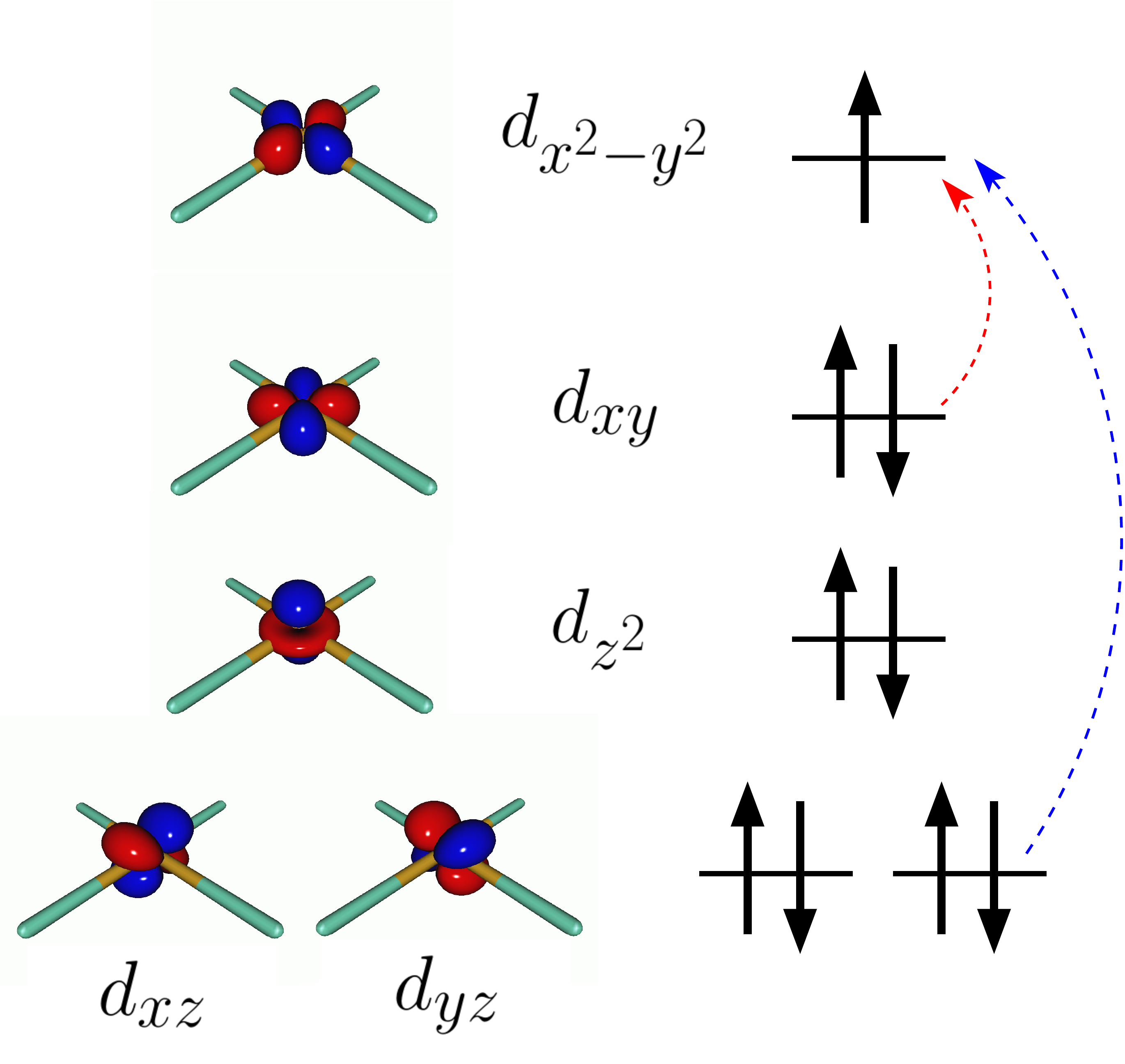}}
\caption{Crystal-field theoretical ordering of the orbitals and the orbital occupation characteristics of the ground state and 
first ({\color{red}{red arrow}}) and third ({\color{blue}{blue arrow}}) lowest $d-d$  electronic transitions.}
\label{qualitative_fig} 
\end{figure}

The three complexes $\cucl$, $\cunh$ and $\cuwater$ are all square-planar coordinated and have a doublet ground state with a
$3d_{x^2 - y^2}$ singly-occupied molecular orbital (SOMO), which has the largest repulsion with the ligand lone-pairs that
point at the Cu atom along the $x$ and $y$ axes. 
Two of the three complexes, $\cucl$ and $\cunh$, have been studied extensively
and the EPR spectra, spin density, $g$-tensor and electronic excitation energies are well characterised experimentally\cite{Morosin-cunh3_4,Hathaway-cunh3_4,experimental_cucl4_1,experimental_cucl4_2}.
The schematic ligand-field diagram\cite{ligand_field} is displayed in
Fig.~\ref{qualitative_fig}. The low-lying excited states all correspond to doublets where one of 
the more low lying $d$ orbitals becomes the SOMO. 

In multi-reference computational studies of two of these systems\cite{solomon_cucl4,neese_CuNH3_SOS,atanasov_cucl4,Remenyi_cucl4,g_tensor_pierloot_mscaspt2}
Neese \textit{et al.} and Pierloot \textit{et al.} observed that in order to correctly describe the electronic spectrum and magnetic properties
it is necessary to include the ligand donor orbital in the active space, even though this orbital 
is doubly occupied and has a relatively low orbital eigenvalue. They also found that CASPT2 performs poorly and
sophisticated methods such as SORCI\cite{neese_CuNH3_SOS} or MS-CASPT2\cite{g_tensor_pierloot_mscaspt2} are required,
which couple the dynamic correlation into the multi-reference treatment.
The general importance of ligand donor orbitals was highlighted by
Nieuwpoort, Broer and coworkers in their pioneering work on cluster models of transition metal 
oxides,\cite{PhysRevB_jansen,broer_1988,vanOosten1996_noci,Broer_noci2002}
where they showed that ligand-to-metal charge transfer (LMCT), and associated orbital relaxation, forms a  
significant component of the wavefunction.
Many subsequent studies have confirmed the importance of LMCT in a range of transition metal
systems,\cite{ddci,carmen_jp_2000_tab,carmen_jp_2002_deloc_nos,malrieu_magnet_3,angeli_carmen_nevpt_spin_1}
and recent work by one of us\cite{cucl2_1,cucl2_cele,fobo_scf} found
analogous correlation mechanisms in several open-shell systems, both inorganic and organic.
A common observation in all of these studies is that the extent of metal-ligand delocalization can 
increase considerably as higher-order correlation effects are taken into account, and the question of what
level of theory is required remains open.

In this work we provide a detailed analysis of wavefunctions, and the role of LMCT in $d-d$ excitation energies and spin-densities of the
$\cucl$, $\cunh$ and $\cuwater$ complexes, and the extent to which these processes are captured in commonly applied wavefunction methods.
The paper is organized as follows. Section \ref{sec_bench} presents benchmark near FCI calculations on the three lowest energy
states of $\cucl$, $\cuwater$ and $\cunh$ and an analysis of the wavefunctions and LMCT in section \ref{section_pt}. 
In section \ref{sec_multi} we discuss the
performance of multi-reference methods for these systems, addressing the appropriate minimal active space required to capture 
the dominant physical processes at play. The performance of single reference methods is discussed in section \ref{sec_single}, 
where we demonstrate that non-size-extensivity errors severvely degrade theoretical predictions, even for these small molecules.
In section \ref{sec_references}, near basis set limit reference transition energies are reported. Our conclusions are summarised in
section \ref{sec_conclu}.

\section{Benchmark near FCI energies and wave functions}
\label{sec_bench}

Near FCI wavefunctions for the ground, first and third electronic states of each of the three complexes were computed using the
CIPSI method in a 6-31G basis set. $D_{2h}$ symmetry ($D_2$ for $\cunh$) was used and 
each state is the lowest energy state in the symmetry block to which it belongs.
He, Ne and Ar cores were frozen in the nitrogen, chlorine and copper atoms, respectively,
resulting in 41 electrons in 50, 66 and 74 orbitals for the $\cucl$, $\cuwater$ and $\cunh$ molecules, respectively.
ROHF orbitals were used to ease comparison of the wavefunction parameters with those of CASSCF, targeted CI and CC-based wavefunctions. 
The geometries of $\cucl$ and $\cunh$ were taken from Ref. \cite{g_tensor_pierloot_mscaspt2}. The geometry of $\cuwater$ was optimized 
with $D_{2h}$ symmetry at the unrestricted PBE\cite{pbe0} level of theory using a 6-31G* basis set. 

The CIPSI approach approximates the FCI energy through an adaptively refined selected CI procedure, 
corrected for discarded determinants through second-order multireference perturbation theory. The CIPSI class of methods build upon 
selected CI ideas\cite{bender,malrieu,buenker1,buenker-book,three_class_CIPSI,harrison,hbci}
and have been successfully used to converge to FCI correlation energies, one-body properties and nodal surfaces.\cite{three_class_CIPSI,Rubio198698,cimiraglia_cipsi,cele_cipsi_zeroth_order,Angeli2000472,canadian,atoms_3d,f2_dmc,atoms_dmc_julien}
The CIPSI algorithm used in this work uses iteratively enlarged selected CI and
Epstein--Nesbet\cite{epstein,nesbet} multi-reference perturbation theory. The CIPSI energy is
\begin{align}
  E_\mathrm{CIPSI} &= E_v + E^{(2)} \\
  E_{v} &= \min_{\{ c_{\rm I}\}} \frac{\elemm{\Psi^{(0)}}{H}{\Psi^{(0)}}  }{\ovrlp{\Psi^{(0)}}{\Psi^{(0)}}} \\
  E^{(2)} &= \sum_{\mu} \frac{\left|\elemm{\Psi^{(0)}}{H}{\mu}\right|^2}{E_{v} - \elemm{\mu}{H}{\mu}} = \sum_{\mu} \,\, e_{\mu}^{(2)} \\
  \ket{\Psi^{(0)}} &= \sum_{{\rm I}\,\in\,\mathcal{R}} \,\,c_{\rm I} \,\,\ket{\rm I}
\end{align}
where I denotes determinants within the CI reference space $\mathcal{R}$ and $\mu$ a determinant outside it.
To reduce the cost of evaluating the second-order energy correction, the semi-stochastic multi-reference approach
of Garniron \textit{et al} \cite{stochastic_pt_yan} was used, adopting the technical specifications recommended in that work. 
The CIPSI energy is systematically refined by doubling the size of the CI reference space at each iteration, selecting
the determinants $\mu$ with the largest $\vert e_{\mu}^{(2)} \vert$, and the energy monitored as a function of the
size of the reference space. 

\subsection{Reference near FCI energies}

\begin{figure}[h]
\subfloat[]{\includegraphics[scale=0.40]{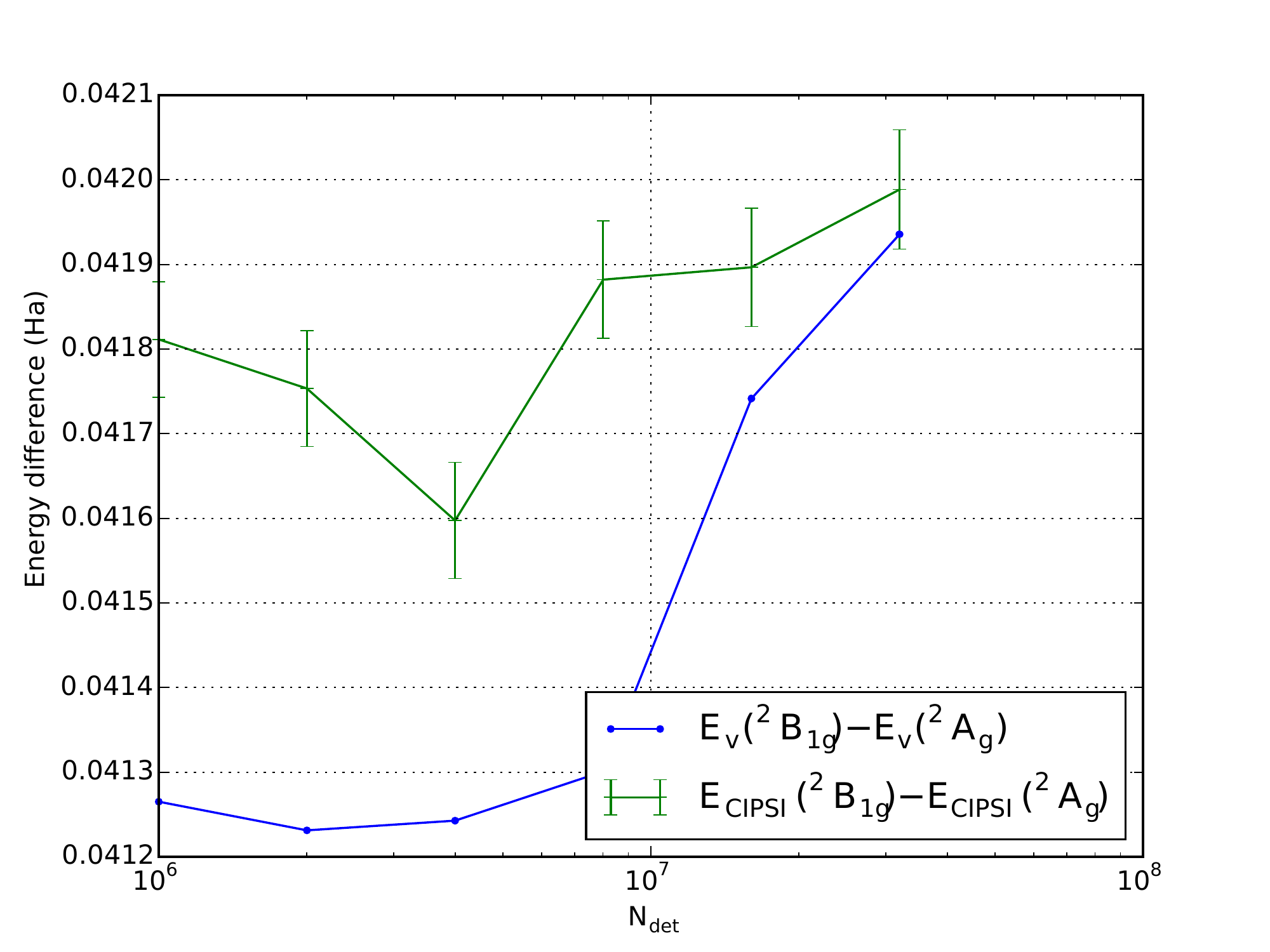}}
\subfloat[]{\includegraphics[scale=0.40]{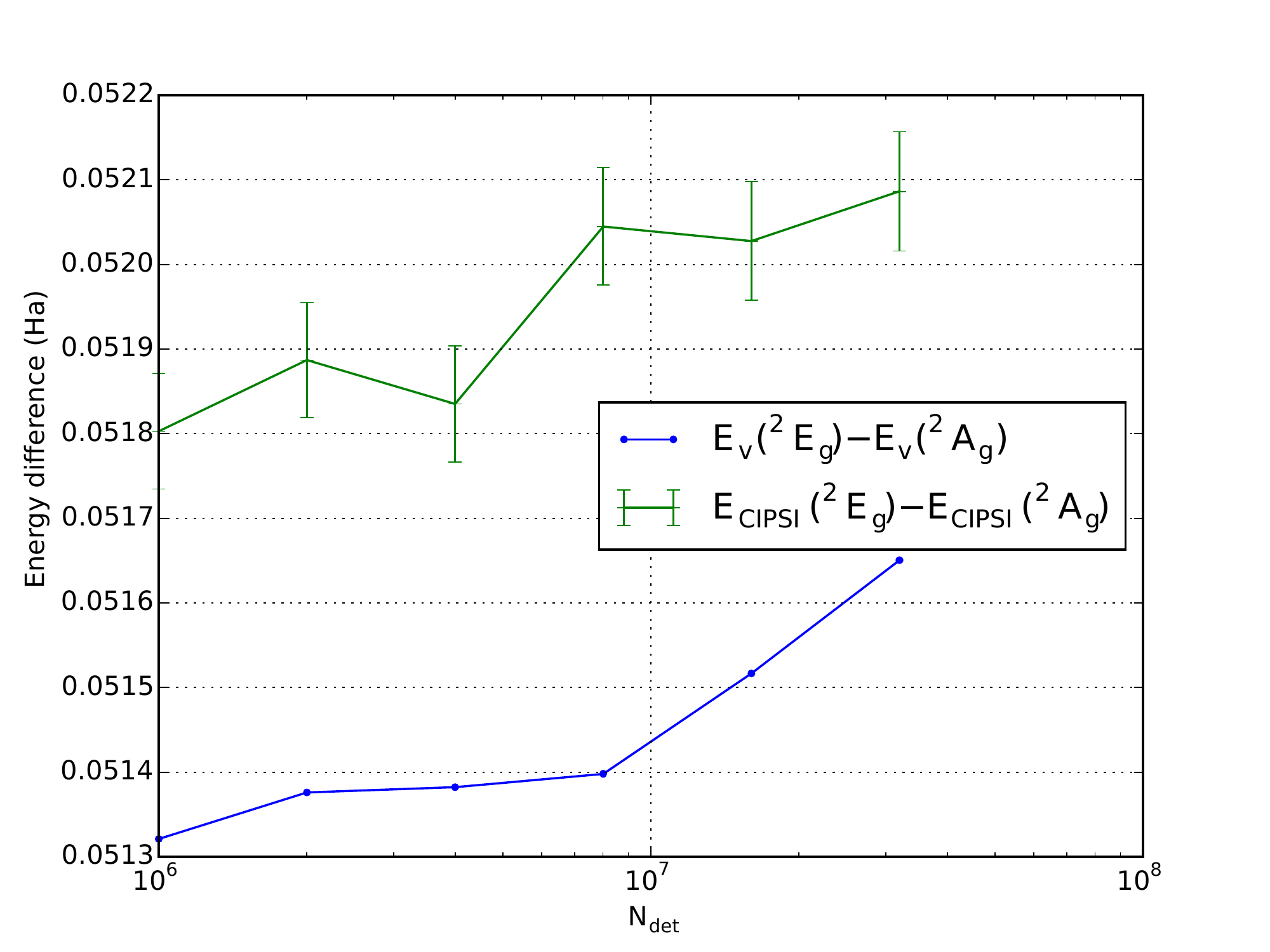}}\\
\subfloat[]{\includegraphics[scale=0.40]{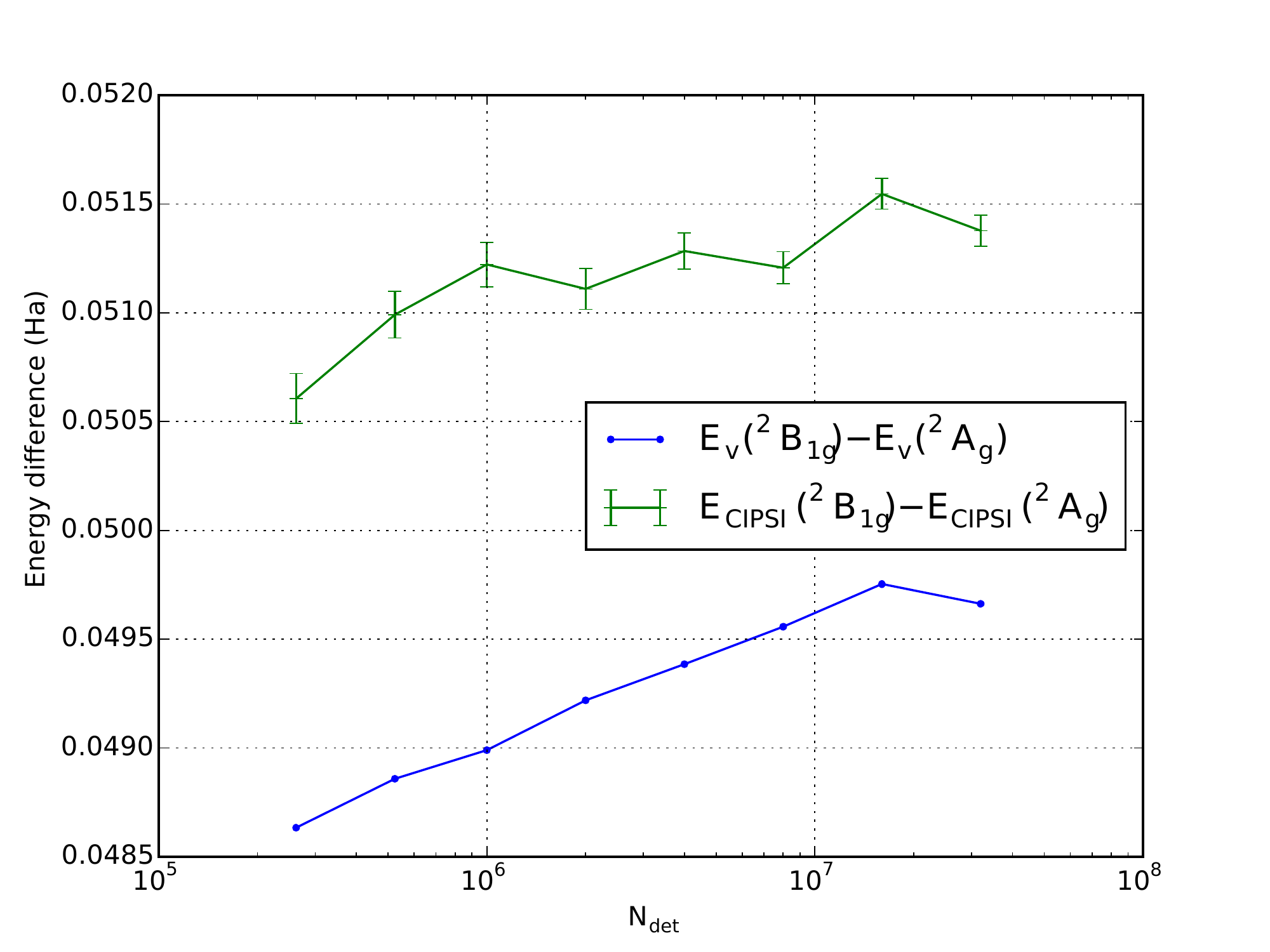}}
\subfloat[]{\includegraphics[scale=0.40]{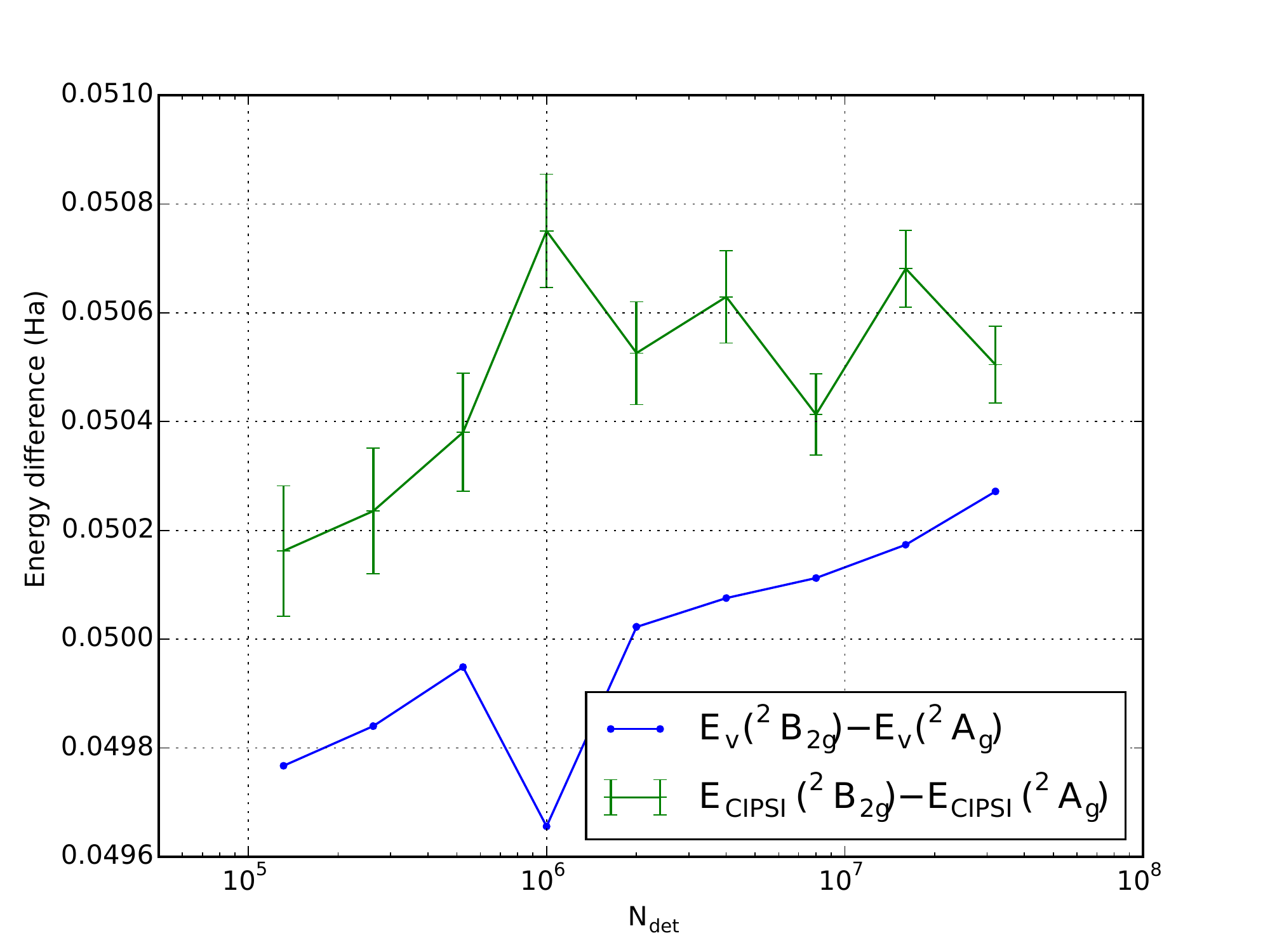}}\\
\subfloat[]{\includegraphics[scale=0.40]{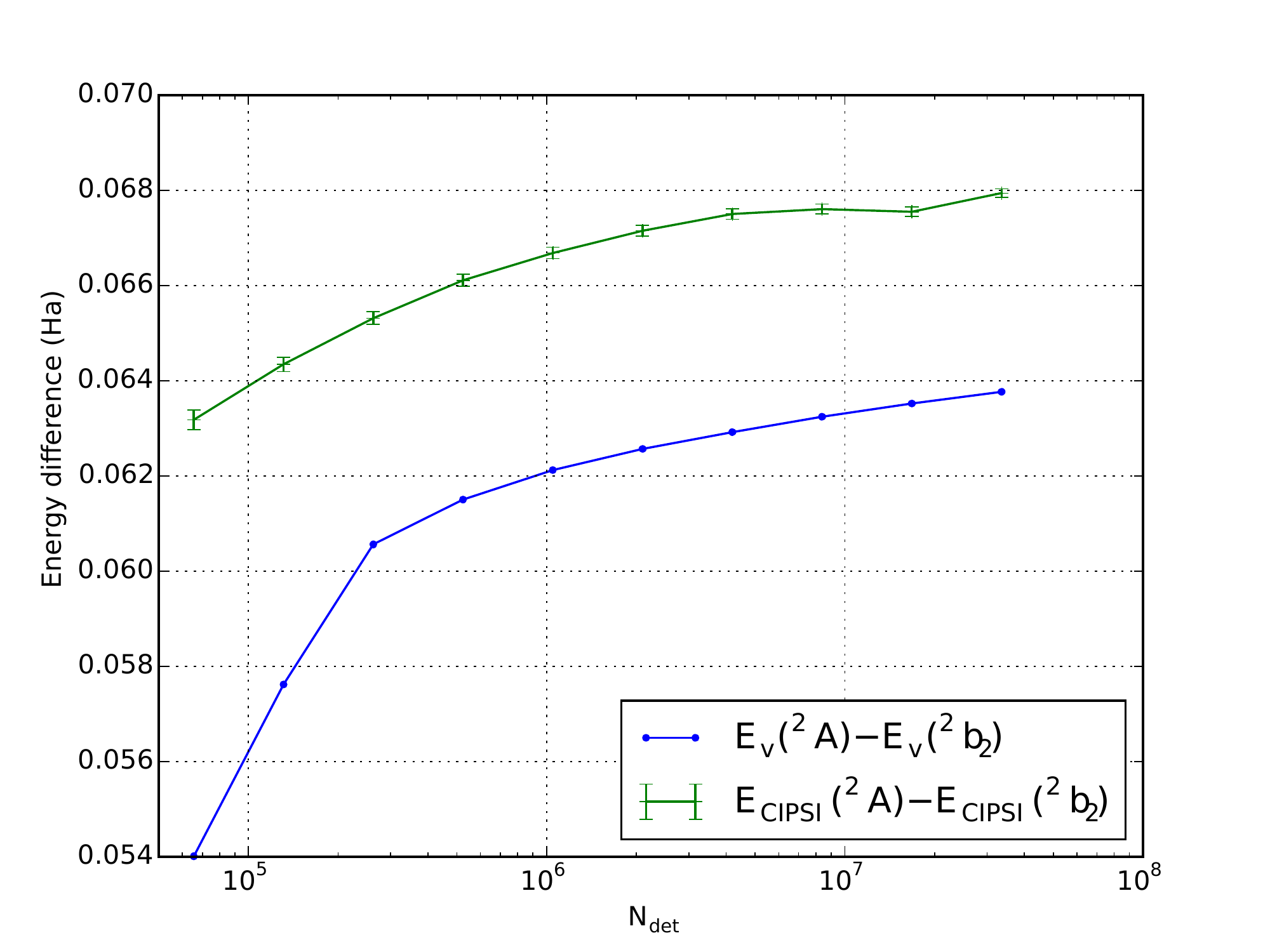}}
\subfloat[]{\includegraphics[scale=0.40]{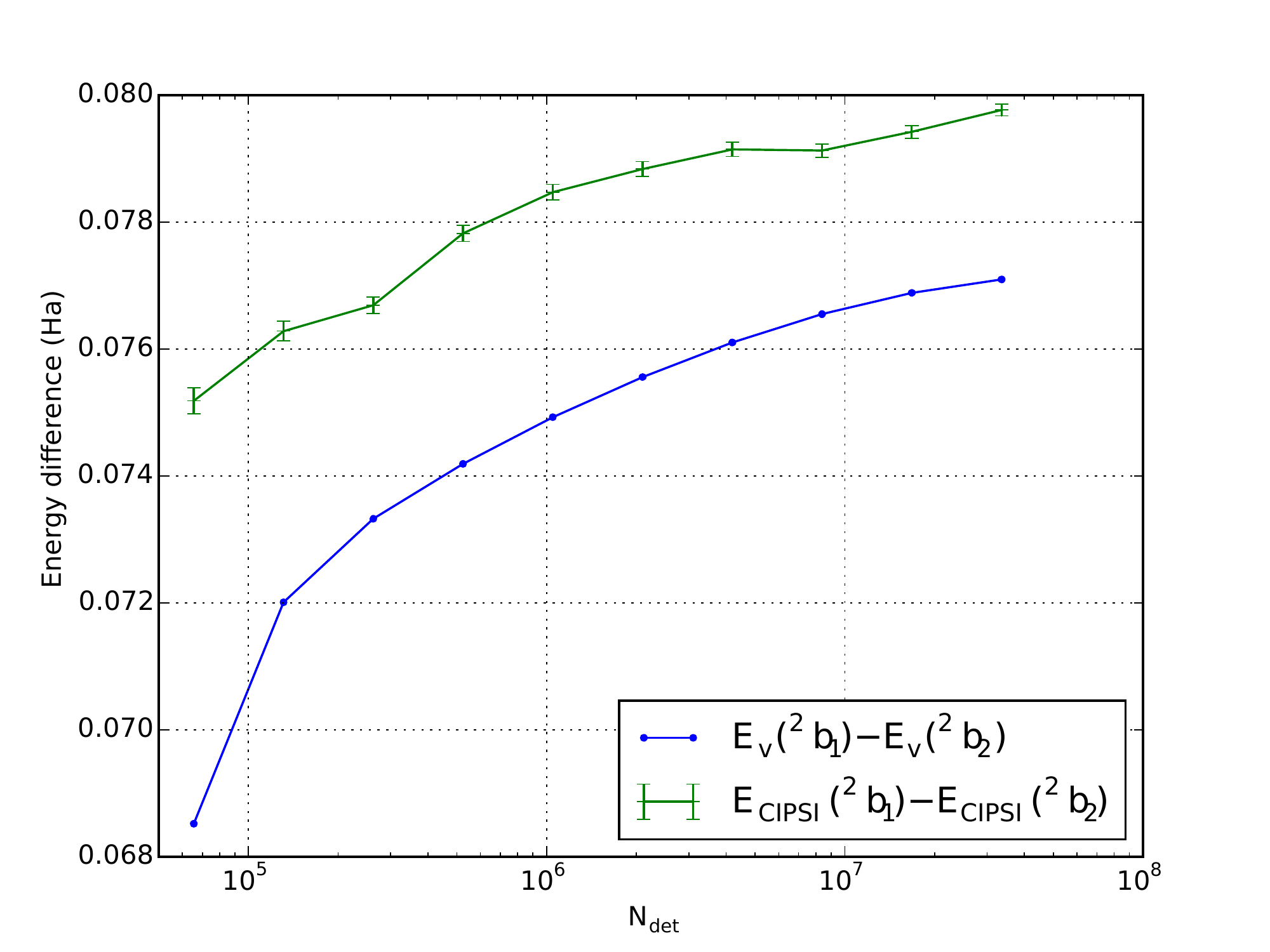}}
\caption{Convergence of the $^2B_{1g}$ $-$ $^2B_{2g}$ and $^2E_g$ $-$ $^2B_{2g}$ 
electronic transitions at the variational and CIPSI level in the 6-31G basis set 
for the $\cucl$ (a and b) and $\cuwater$ (c and d) complex and of the 
$^2B_1$ $-$ $^2B_2$  and $^2E$ $-$ $^2B_2$ electronic transition (e and f) 
for the $\cunh$ complex as a function of the size of the reference CIPSI wave function. \label{conv_cipsi} }
\end{figure}

Fig. \ref{conv_cipsi} plots convergence with respect to size of the CIPSI reference wave function for the electronic transitions 
of the $\cucl$ $\cuwater$ and $\cunh$ complexes, up to 32$\times 10^6$ Slater determinants for the $\cucl$ and $\cuwater$ systems and
64$\times 10^6$ Slater determinants for the $\cunh$ molecule. Both $E_v$ and $E_\mathrm{CIPSI}$ are displayed.
Here and throughout $D_{4h}$ and $D_{2d}$ symmetry labels are used for the electronic states.
The CIPSI electronic transition energies for the $\cucl$ and $\cuwater$ molecules are converged with a sub-mH precision within 8$\times 10^6$ Slater determinants and the variational CI transition energies agree with the CIPCI values to within 1~mH.
Due to the larger Hilbert space, the convergence for the $\cunh$ molecule is significantly slower. The variational CI energy difference is not converged even using 64$\times 10^6$ Slater determinants, but the CIPSI values do appear converged to within 1~mH with 64$\times10^6$ Slater determinants, underlining the importance of the second order correction to the energy. 
We note that in all cases there is a clear trend: increasing the CI reference space increases the energy differences, which indicates that the
ground state has a larger correlation energy than that of the excited states for each of the molecules.

\begin{table}
\begin{center}
\caption{Computed excitation energies (mH) for the $\cucl$, $\cunh$ and $\cuwater$ molecules. }
\scalebox{0.7}{
\begin{tabular}{|l|ccccccc|ccc|c|}
\hline
Electronic transition      & ROHF  & CISD  & CISD(SC)$^2$& CCSD      & BCCD    &  CCSD(T)  &BCCD(T)& CAS(9-10) &CAS(11-11) & FOBOCI     & CIPSI     \\
\hline                                                                                                                                                                       
\multicolumn{12}{|c|}{$\cucl$}\\                                                                                                                  
\hline                                                                                                                                                                       
$^2B_{1g}$ $-$ $^2B_{2g}$    & 30.2  & 38.3  & 43.6        & 43.2      & 42.6    & 43.9      & 43.8  &  31.1     &  39.7     & 43.2       &  42.0(1)   \\
$^2E_{g}$  $-$ $^2B_{2g}$    & 38.2  & 46.7  & 52.2        & 51.9      & 51.3    & 52.6      & 52.6  &  39.9     &  48.8     & 51.4       &  52.1(2)   \\
\hline                                                                                                                                                                       
\multicolumn{12}{|c|}{$\cuwater$}\\                                                                                                               
\hline                                                                                                                                                                       
$^2B_{1g}$ $-$ $^2B_{2g}$    & 42.4  & 48.1  & 51.6        &  51.6     &  51.3   & 52.1      & 52.1  &  44.2     &  48.9     & 49.9       &  51.5(1)   \\
$^2E_{g}$  $-$ $^2B_{2g}$    & 44.8  & 48.5  & 50.5        &  50.7     &  50.6   & 50.9      & 50.9  &  46.7     &  50.8     & 50.7       &  50.5(1)   \\
\hline                                                                                                                                                                       
\multicolumn{12}{|c|}{$\cunh$}\\                                                                                                                   
\hline                                                                                                                                                                       
$^2B_{1}$ $-$ $^2B_{2}$      & 50.7  & 60.2  & 68.9        &  68.5     & 67.8    &  69.7     &  69.8 &  53.0     &  62.6     & 66.0       &  68.0(1)   \\ 
$^2E$     $-$ $^2B_{2}$      & 62.7  & 72.1  & 80.5        &  80.5     & 79.8    &  81.6     &  81.7 &  65.6     &  75.3     & 78.3       &  79.9(1)   \\ 
\hline
\end{tabular}
\label{results_cu}
}
\end{center}
\end{table}

\subsection{Composition of the ground state wave functions}
\begin{figure}[h]
\subfloat[SOMO]{\includegraphics[scale=0.25]{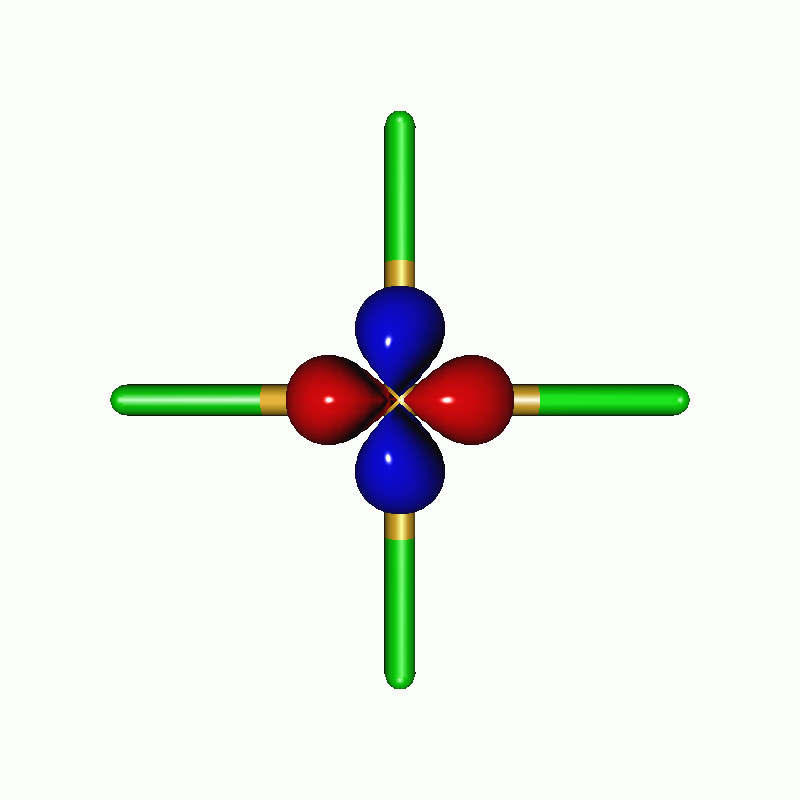}}
\subfloat[Donor orbital]{\includegraphics[scale=0.25]{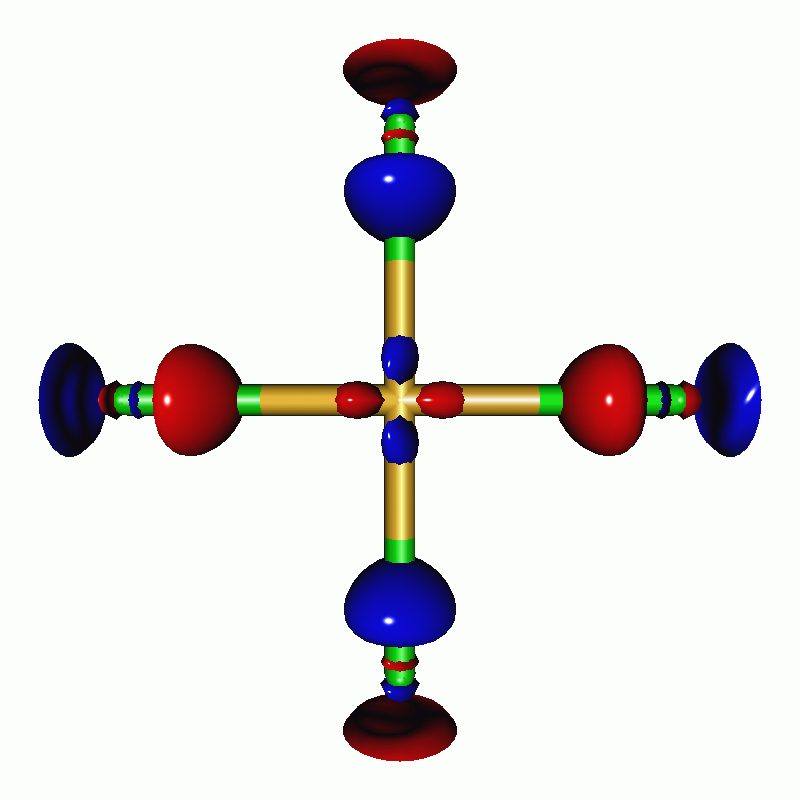}}
\caption{SOMO $S$ at the ROHF level (a) and ligand donor orbital $L$ (b) in the  $^2B_{2g}$ ground state of the $\cucl$ molecule. }
\label{orb_cucl4}
\end{figure}

\begin{figure}[h]
\subfloat[SOMO]{\includegraphics[scale=0.25]{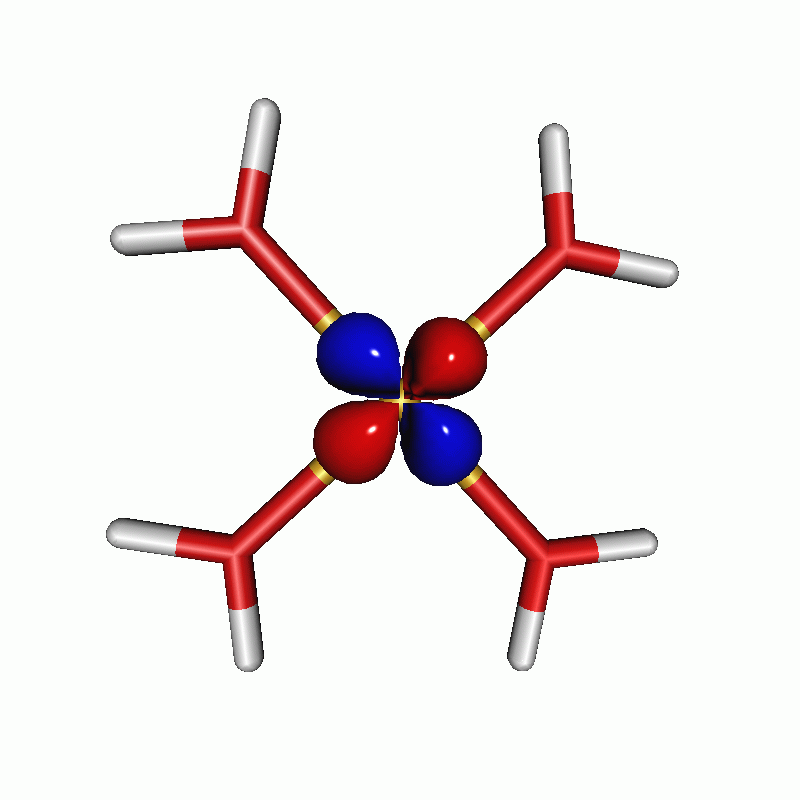}}
\subfloat[Donor orbital]{\includegraphics[scale=0.25]{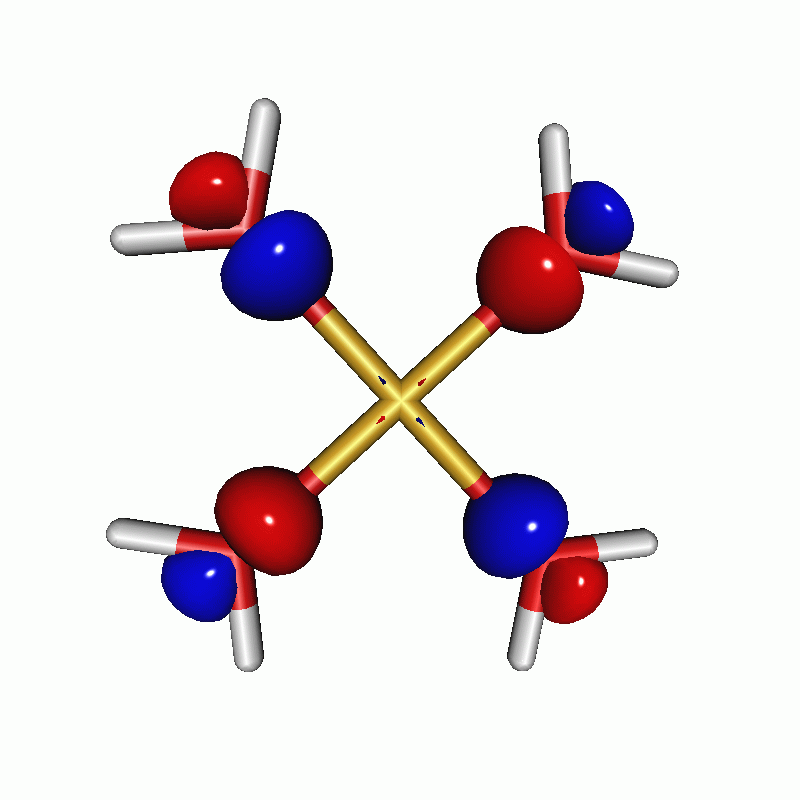}}
\caption{SOMO $S$ at the ROHF level (a) and ligand donor orbital $L$ (b) in the $^2B_{2g}$ ground state of the $\cuwater$ molecule. }
\label{orb_cuwater}
\end{figure}

\begin{figure}[h]
\subfloat[SOMO]{\includegraphics[scale=0.25]{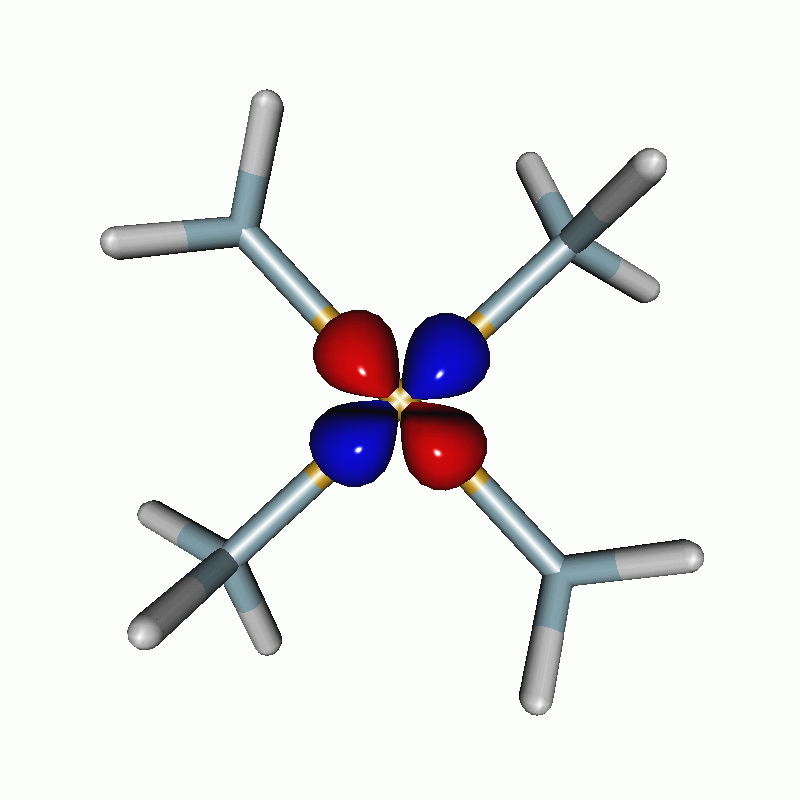}}
\subfloat[Donor orbital]{\includegraphics[scale=0.25]{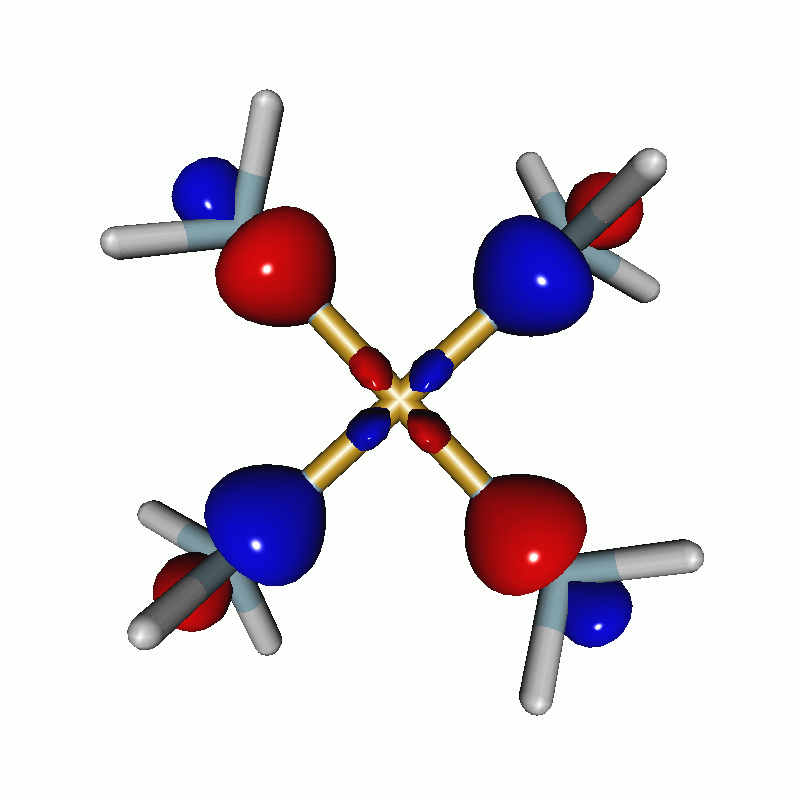}}
\caption{SOMO $S$ at the ROHF level (a) and ligand donor orbital $L$ (b) in the $^2B_2$ ground state of the $\cunh$ molecule. }
\label{orb_cunh3}
\end{figure}

The composition of the near FCI ground state wave functions on the $\cucl$, $\cuwater$ and $\cunh$ complexes present strong similarities in their dominant components: in all cases there are clearly two leading Slater determinants, the ROHF determinant and a single excitation where an electron has been excited from a doubly occupied ligand-based MO to the $3d_{x^2-y^2}$ SOMO on the copper centre. Table \ref{table_cu} lists the amplitudes of these single excitations,
extracted from the largest CIPSI wave function. These singly excited determinants are identified as
a LMCT component of the ground state wavefunction and the orbitals involved are plotted in Figs \ref{orb_cucl4}, \ref{orb_cuwater} and \ref{orb_cunh3}.
Since the ROHF orbitals are reasonably well localized on the Cu atom or on the ligands, one can analyze the physical content of the CIPSI wave functions in terms of valence bond (VB) structures. 
In all three ground states the ROHF determinant corresponds to a VB form of the type $\text{Cu}^{2+}\text{X}_4$
and the LMCT components correspond to a set of four equivalent VB structures of the type $\text{Cu}^{+}\text{X}^+ \text{X}_3$, where X denotes the ligand.
In the ROHF wavefunction, the spin density is concentrated at the copper atom, whereas in the FCI wavefunction, 
the LMCT excitations delocalise the spin-density onto the ligands. The spin-densities are listed in Table \ref{results_cu_sd}.

\begin{table}
\begin{center}
\caption{Coefficient of the largest single excitations at various levels of theory.}
\begin{tabular}{|l|ccccc|}
\hline
  Electronic State            &   CISD     &CISD(SC)$^2$ &  CCSD       & FOBOCI &   CIPSI           \\ 
\hline                                                                                                                                                                        
\multicolumn{6}{|l|}{$\cucl$}\\
\hline                                                                                                                                                                        
$^2B_{2g}$                    &   0.071    &  0.175      &   0.165     &  0.149 &      0.156        \\
$^2B_{1g}$                    &   0.032    &  0.083      &   0.078     &  0.064 &      0.069        \\
$^2E_{g}$                     &   0.031    &  0.089      &   0.085     &  0.078 &      0.074        \\
\hline                                                                                                                     
\multicolumn{6}{|l|}{$\cuwater$}\\
\hline                                                                                                                     
$^2B_{2g}$                    &   0.028    &  0.076      &   0.075     & 0.065  &      0.060        \\
$^2B_{1g}$                    &   0.001    &  0.028      &   0.028     & 0.012  &      0.006        \\
$^2E_{g}$                     &   0.005    &  0.058      &   0.056     & 0.050  &      0.035        \\
\hline                                                                                                                     
\multicolumn{6}{|l|}{$\cunh$}\\
\hline                                                                                                                    
$^2B_2$                       &   0.043    &  0.141      &   0.137     & 0.117  &      0.113        \\ 
$^2B_1$                       &   0.001    &  0.015      &   0.033     & 0.016  &      0.009        \\ 
$^2E$                         &   0.001    &  0.014      &   0.028     & 0.014  &      0.006        \\ 

\hline
\end{tabular}
\label{table_cu}
\end{center}
\end{table}

\begin{table}
\caption{Spin density on the copper atom at various levels of theory using the Mulliken population analysis. }
\resizebox{\columnwidth}{!}{
\begin{tabular}{|l|cccc|ccc|c|}
\hline
Electronic state                & ROHF &CISD(SC)$^2$&  CISD  & BCCD        & CAS(9-10) &CAS(11-11) & FOBOCI     & CIPSI     \\
\hline                                                                                                                                                  
\multicolumn{9}{|c|}{$\cucl$}\\                                                                                                
\hline                                                                                                                                                  
$^2B_{2g}$                      & 0.93 & 0.80       & 0.89   &  0.80       & 0.93      &   0.86    &  0.81      &  0.81      \\
$^2B_{1g}$                      & 0.99 & 0.99       & 1.00   &  1.00       & 0.99      &   0.98    &  0.99      &  0.99      \\
$^2E_{g}$                       & 0.99 & 0.99       & 1.00   &  0.99       & 0.99      &   0.98    &  0.99      &  0.99      \\
\hline                                                                                                                                                  
\multicolumn{9}{|c|}{$\cuwater$}\\                                                                                                
\hline                                                                                                                                                  
$^2B_{2g}$                      & 0.96 & 0.91       & 0.94   & 0.91        & 0.95      &   0.85    &  0.91      &  0.92      \\
$^2B_{1g}$                      & 0.99 & 1.00       & 1.00   & 1.00        & 0.99      &   0.98    &  1.00      &  1.00      \\
$^2E_{g}$                       & 0.99 & 0.99       & 1.00   & 0.99        & 0.99      &   0.96    &  0.99      &  0.99      \\
\hline                                                                                                                                                  
\multicolumn{9}{|c|}{$\cunh$}\\                                                                                                
\hline                                                                                                                                                  
$^2B_2$                         &  0.92& 0.83       & 0.90    & 0.81       & 0.93       &  0.87     & 0.82       &  0.84      \\ 
$^2B_1$                         &  0.99& 1.02       & 1.00    & 1.02       & 0.99       &  0.99     & 1.01       &  1.01      \\ 
$^2E$                           &  0.99& 1.02       & 1.00    & 1.02       & 0.99       &  0.99     & 1.01       &  1.01      \\ 
\hline
\end{tabular}
\label{results_cu_sd}
}

\end{table}

\subsection{Composition of the excited state wave functions}
The composition of the CIPSI wave function for the various excited states presents strong similarities with the ground state wave functions. 
In all cases, LMCT single excitations appear where an electron is transfered from a doubly occupied ligand orbital with the same symmetry
as the SOMO, to the SOMO at the copper centre. The orbitals involved in the LMCT processes are displayed in 
Figs. \ref{orb_cuwater_e1}, \ref{orb_cunh_e1} and \ref{orb_cunh_e2} and the amplitudes are reported in Table \ref{table_cu}, 
as extracted from the largest CIPSI wavefunctions. The LMCT excitations are evidently important in the $^2B_{1g}$ and $^2E_{g}$ excited states 
of $\cucl$ and in the $^2E_{g}$ state $\cuwater$, but with an amplitude half the magnitude of that of the ground state, due to the weaker overlap
of the ligand and metal orbitals. The amplitudes of the LMCT excitations for the $^2B_{1g}$ state of $\cuwater$ and both states of 
$\cunh$ are between one and two orders of magnitude smaller.

\section{Perturbation theory analysis of LMCT}
\label{section_pt}

Our benchmark near FCI wavefunctions and energies in the 6-31G basis set reveal that the
ground state is more correlated than the excited states in all three complexes, and that 
the LMCT processes are stronger in the ground state than in the excited states. 
In this section we analyse in greater depth the role of electron correlation and LMCT 
in the transition energies from the perspective of single reference perturbation theory.
Here and throughout, all the orbitals doubly occupied in the ROHF Slater determinant are referred as $i$, the ligand \textit{donor} 
orbital for each state is called $L$, the SOMO as $S$ and the virtual orbitals as $a$. 
Also, the $S_z$ component is assumed to be $\frac{1}{2}$ so the unpaired electron has $\alpha$ spin. The LMCT determinant is
\begin{equation}
 \ket{\rm LMCT} = a^{\dagger}_{S\beta} a_{L\beta} \ket{\rm ROHF}
\end{equation}
The prevalence of the LMCT determinants in all ground and some excited states wavefunctions can be considered as quite unusual at least for two reasons. First, all LMCT determinants are more than 12 eV higher in energy than the ROHF determinant, which is clearly not a near degeneracy situation.  
Second, coefficient of the LMCT determinant at first-order in M\o ller--Plesset perturbation (MP) theory\cite{mp} is 
\begin{equation}
 c^{(1)}_{\rm LMCT} = \frac{\elemm{\rm LMCT}{H}{\rm ROHF}}{E^{(0)}_0 - E^{(0)}_{\rm LMCT}} = 0
\end{equation}
which vanishes because of the Brillouin Theorem. 
The large coefficients of the LMCT determinant in the near FCI wave function come necessarily from their interactions with 
determinants of higher excitation rank.
The first non-vanishing contribution to the LMCT coefficient appears at second order in the MP expansion:
\begin{equation}
 c_{\rm LMCT}^{(2)} = \sum_{\rm D} \frac{\elemm{\rm LMCT}{H}{\rm D}}{E^{(0)}_0 - E^{(0)}_{\rm LMCT}} c^{(1)}_{\rm D}  = \sum_{\rm D} \frac{\elemm{\rm LMCT}{H}{\rm D}}{E^{(0)}_0 - E^{(0)}_{\rm LMCT}} \frac{\elemm{\rm D}{H}{\rm ROHF}}{E^{(0)}_0 - E^{(0)}_{\rm D}} \equiv \sum_{D} \delta c_{\rm D}
\end{equation}
The contribution $\delta c_{\rm D}$ of each double excitation $\ket{\rm D}$ to the second-order coefficient can be used
to identify the most important double excitations for the LMCT. The largest values of $|\delta c_{\rm D}|$ correspond 
in all three states of all three molecules to the specific class of double excitations that are single excitations from
the LMCT determinant
\begin{equation}
 \label{double_ct}
 \ket{_L ^{S}\, _{i,\sigma} ^{a,\sigma}} \equiv a^{\dagger}_{a\sigma} a_{i\sigma} a^{\dagger}_{ S\beta} a_{L\beta} \ket{\rm ROHF}=  a^{\dagger}_{a\sigma} a_{i\sigma} \ket{\rm LMCT}
\end{equation}
Among these, the largest $|\delta c_{\rm D}|$ occur when $i$ is a $3d$ and $a$ a $4d$ orbital, where the
the interaction elements $\elemm{\rm LMCT}{H}{_L ^{S}\, _i ^a}$ are found to be around 7 eV, which is 
very large for off-diagonal Hamiltonian matrix elements. 
The large magnitude can be easily understood: applying the Brillouin-Theorem and neglecting minor exchange contributions,
the pertinent matrix elements are
\begin{equation}
\label{f_ia}
 \elemm{\rm LMCT}{H}{_L ^{S}\,  _{i,\sigma} ^{a,\sigma}} \approx (ia|SS) - (ia|LL)
\end{equation}
where the standard chemical notation is used for the two-electron repulstion integrals. The integrals $(ia|SS)$ are very large 
(typically between 7 and 8 eV in our calculations) since all orbitals are located at the copper atom, and the integrals
$(ia|LL)$ are small (typically between 0.1 and 0.5 eV) since the 
distributions $ia$ and $LL$ are centered on different atomic sites.

The fact that the single excitations from \lmct are important can be interpreted physically as the need to relax 
the orbitals of the \lmct determinant. The ROHF orbitals are not optimal for the \lmct determinant since the
former represents the copper atom in its Cu$^{2+}$, wheras the latter represents the copper atom in its Cu$^+$ state where
the orbitals are more diffuse.
This is nothing other than the breathing-orbital effect, well known in the VB framework.\cite{breath_vb}  
These considerations all point to a subtle 
interplay between electronic correlation and metal-ligand delocalization in the spectroscopy
of these transition metal complexes.

\section{Multi-reference methods}
\label{sec_multi}

When wavefunction approaches are applied to transition metal systems, multi-reference (MR) methods are usually selected.
The results obtained often depend critically on the choice of active space and in this section we 
examine the influence of the active space on the transition energies and spin densities.

\subsection{CASSCF}

A common choice of active space in transition metals is the so-called 'double $d$-shell', which involves all valence $3d$ electrons 
in the $3d$ and $4d$ orbital sets. For $3d^{9}$ copper complexes, this results in a CAS(9-10), nine electrons in ten orbitals.
The CASSCF transition energies and spin densities are reported in Tables \ref{results_cu} and \ref{results_cu_sd}, respectively, 
and compared to the corresponding near FCI values. Although this active space captures the dominant dynamical correlation of the  
the $3d$ electrons, it is often insufficient for accurate results and this is also the case here. The computed electronic
transition at the CAS(9-10) level are 8 to 15 mh too low and the spin density on the copper atom is overestimated, with
almost no improvement over ROHF for both quantities.

The analysis in the previous section highlights the importance of LMCT single excitations $3d\rightarrow L$, 
which are missing from the CAS(9-10) active space that contains only $3d\rightarrow4d $ excitations.
Adding the ligand donor orbital $L$ to the active space results in a CAS(11-11) and the CASSCF 
transition energies and spin densities are also reported in Tables \ref{results_cu} and \ref{results_cu_sd}.
The results are substantially improved due to the presence of LMCT single excitations in the active space, but
significant deviations from the reference CIPSI values remain.

\subsection{A minimal CI space: FOBOCI }
The perturbation analysis of Sec. \ref{section_pt} suggests that it is possible to define a minimal selected CI,
refered to as FOBOCI\cite{cucl2_cele,fobo_scf,giner_hcc} 
(first order breathing orbital CI) that contains the dominant physical effects related to the LMCT determinant. 
The minimal CI should contain the ROHF and LMCT determinants, 
together with all single excitations from these two configurations to introduce the necessary orbital relaxation.
The FOBOCI therefore contains all single excitations and all double excitations of type $\ket{_L ^{S}\, _i ^a}$.

The results obtained at the FOBOCI level for the electronic transitions are reported in Table \ref{results_cu}, the amplitude of the LMCT 
determinants in the FOBOCI wave function are reported in Table \ref{table_cu} and the spin density on the copper atom is reported in 
Table \ref{results_cu_sd}. The FOBOCI electronic transition energies are remarkably close to the near FCI values, with a mean error of
1.2 mH and a maximum error of 2 mH. The amplitudes of the LMCT determinants in the FOBOCI wave functions are also 
close to those of the CIPSI wave function, as are the spin densities at the copper atom.
The FOBOCI wavefunction clearly contains the dominant differential physical effects involved in the spectroscopy of these complexes,
correctly balancing electron correlation and spin-delocalisation.

The success of the FOBOCI wavefunctions is even more remarkable considering that they contian least 4 orders of magnitude fewer determinants  
than the CIPSI wave functions: the largest FOBOCI wave function contains 1072 Slater determinants in the case of the $\cunh$ molecule, 
compared to 64 $\times 10^6$ Slater determinants of the CIPSI wave functions. Although the FOBOCI wavefunction only recovers
about 3\% of the total correlation energy of each state, this small fraction of the correlation contribution has a large
differential effect on the energies of the ground and excited states.

It is also interesting to note that CAS(11-11) performs systematically worse than FOBOCI, even though the CAS(11-11) total energies
are $\sim$0.1 Hartree below the FOBOCI values. The main difference between the FOBOCI and the CAS(11-11) wavefunctions 
is that the former contains $\ket{_{Li}^{Sa}}$ determinants with $i$ and $a$ located on the ligands, which are missing from the CAS(11-11)
active space. The $\ket{_{Li}^{Sa}}$ determinants with $i$ and $a$ are $3d$ and $4d$ orbitals allow for the dilatation of the copper orbitals due to the transfer of charge from the ligand to the copper, and the $\ket{_{Li}^{Sa}}$ determinant with $i$ and $a$ located on the ligands allows for the
corresponding relaxation of the ligand orbitals. Both are required for quantitative agreement with the near FCI results.

\subsection{Perturbation analysis of FOBOCI}
\label{mbpt_analysis}
\begin{figure}[h]
\subfloat[SOMO]{\includegraphics[scale=0.25]{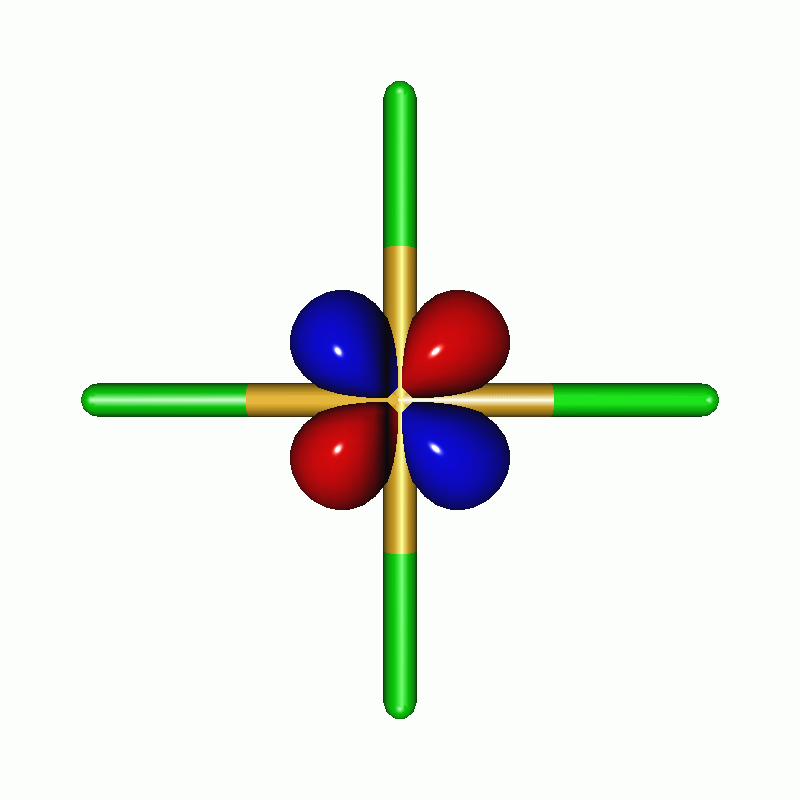}}
\subfloat[Donor orbital]{\includegraphics[scale=0.25]{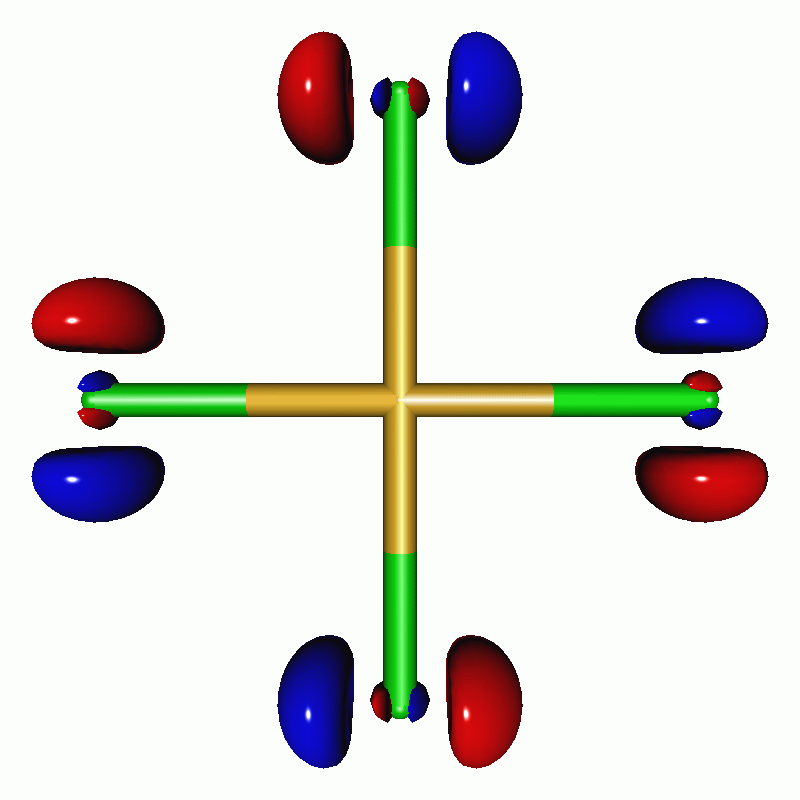}}
\caption{SOMO $S$ at the ROHF level (a) and ligand donor orbital $L$ (b) in the  $^2B_{1g}$ excited state of the $\cucl$ molecule. }
\label{orb_cucl4_e1}
\end{figure}

\begin{figure}[h]
\subfloat[SOMO]{\includegraphics[scale=0.25]{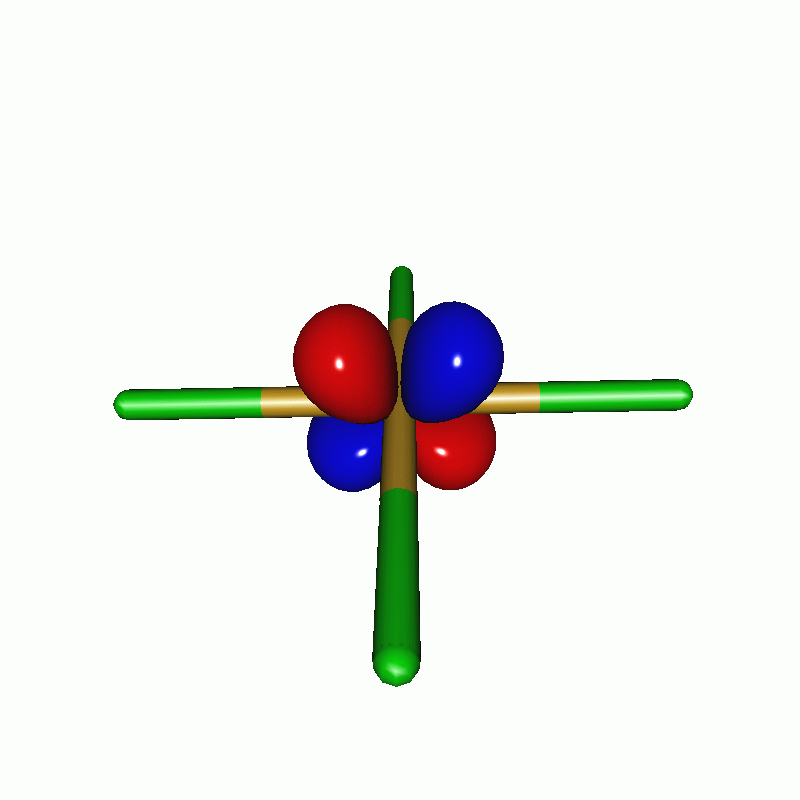}}
\subfloat[Donor orbital]{\includegraphics[scale=0.25]{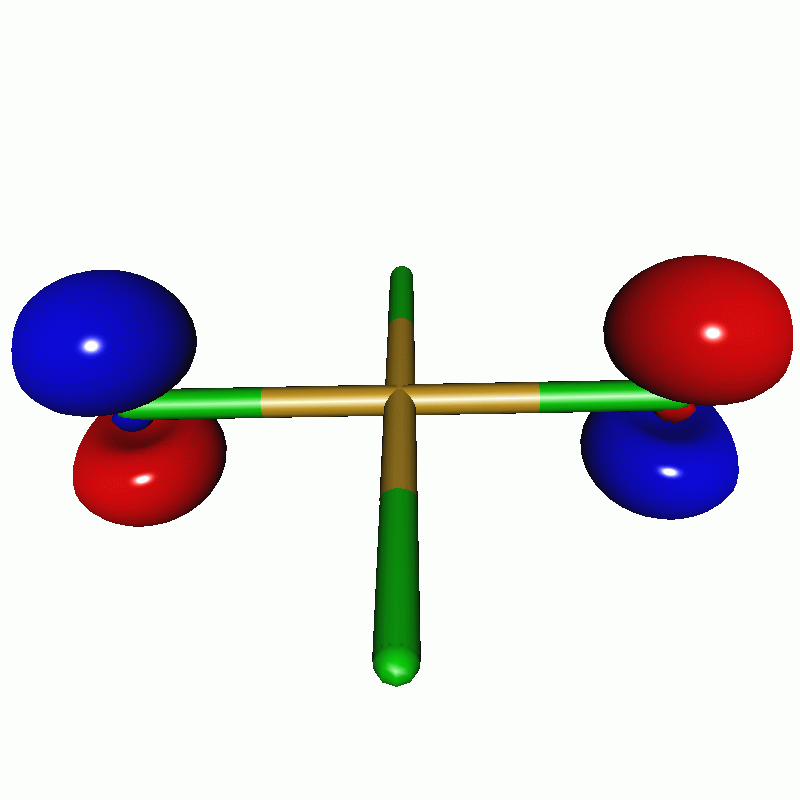}}
\caption{SOMO $S$ at the ROHF level (a) and ligand donor orbital $L$ (b) in the  $^2E_{g}$ excited state of the $\cucl$ molecule. }
\label{orb_cucl4_e2}
\end{figure}

\begin{figure}[h]
\subfloat[SOMO]{\includegraphics[scale=0.25]{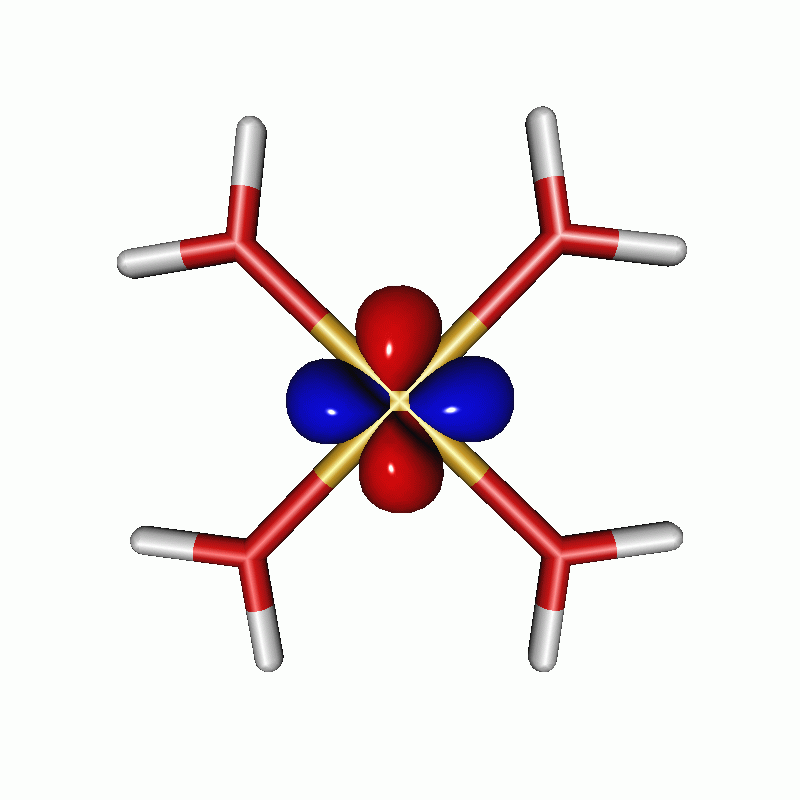}}
\subfloat[Donor orbital]{\includegraphics[scale=0.25]{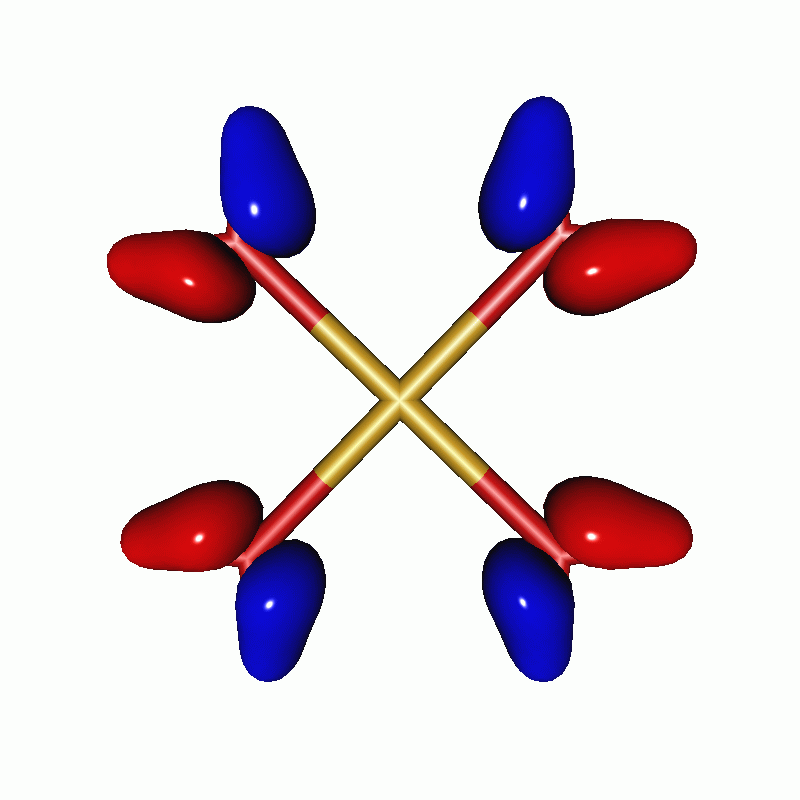}}
\caption{SOMO $S$ at the ROHF level (a) and ligand donor orbital $L$ (b) in the  $^2B_{1g}$ excited state of the $\cuwater$ molecule. }
\label{orb_cuwater_e1}
\end{figure}

\begin{figure}[h]
\subfloat[SOMO]{\includegraphics[scale=0.25]{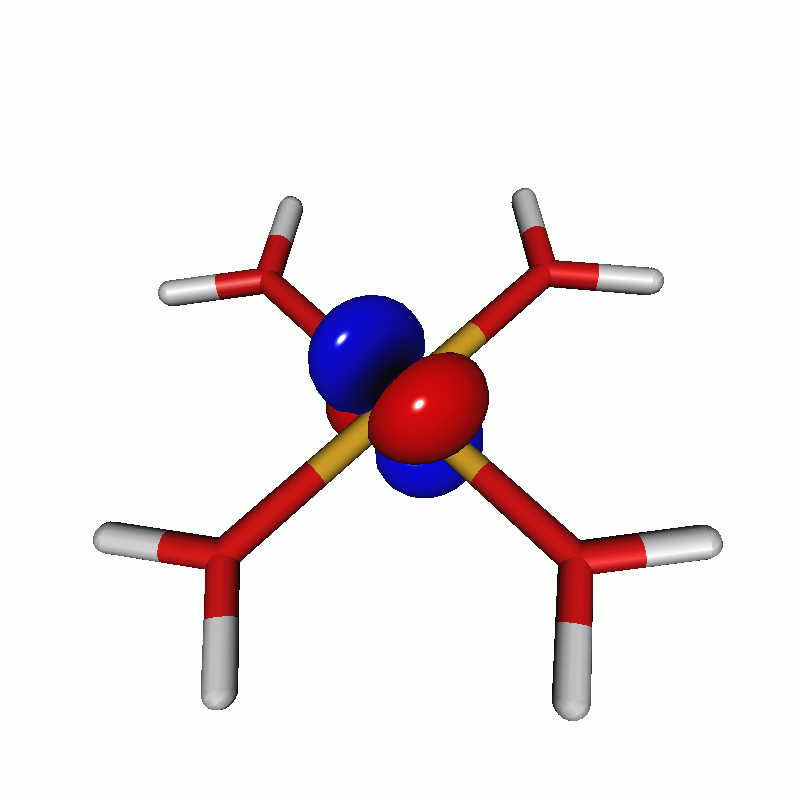}}
\subfloat[Donor orbital]{\includegraphics[scale=0.25]{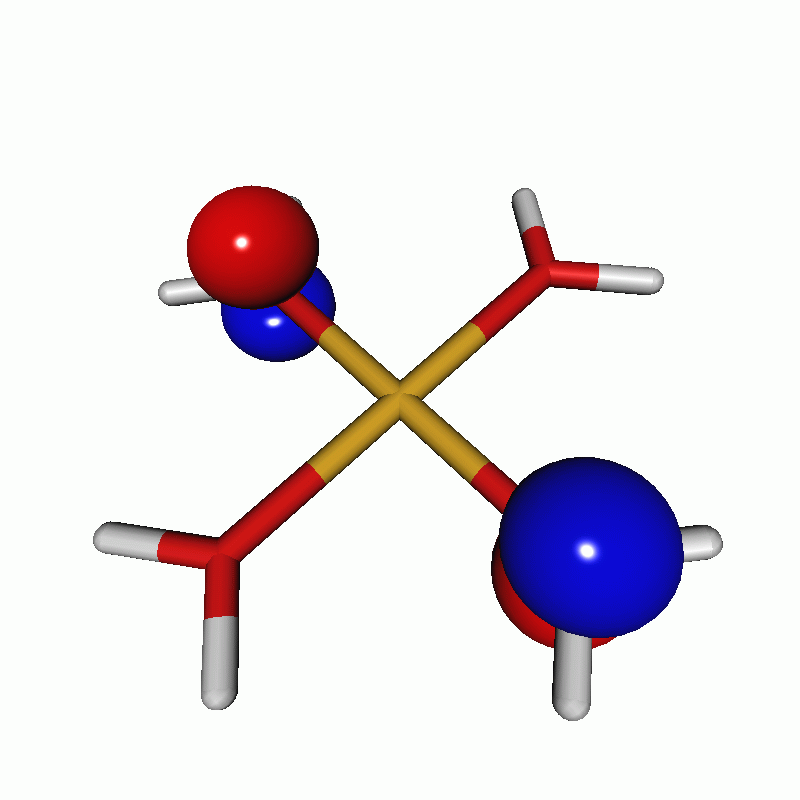}}
\caption{SOMO $S$ at the ROHF level (a) and ligand donor orbital $L$ (b) in the  $^2E_{g}$ excited state of the $\cuwater$ molecule. }
\label{orb_cuwater_e2}
\end{figure}

\begin{figure}[h]
\subfloat[SOMO]{\includegraphics[scale=0.25]{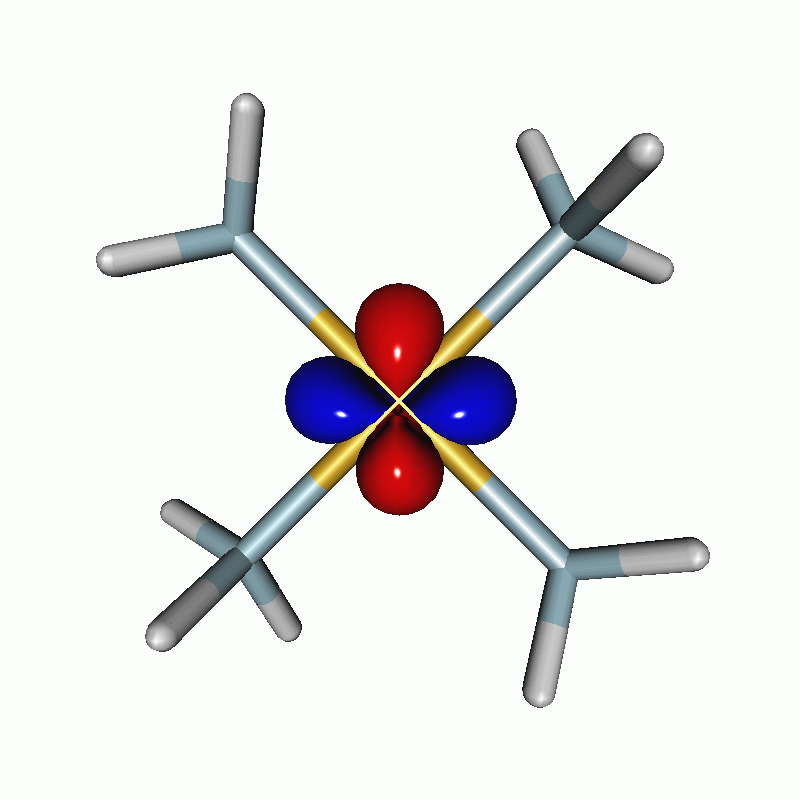}}
\subfloat[Donor orbital]{\includegraphics[scale=0.25]{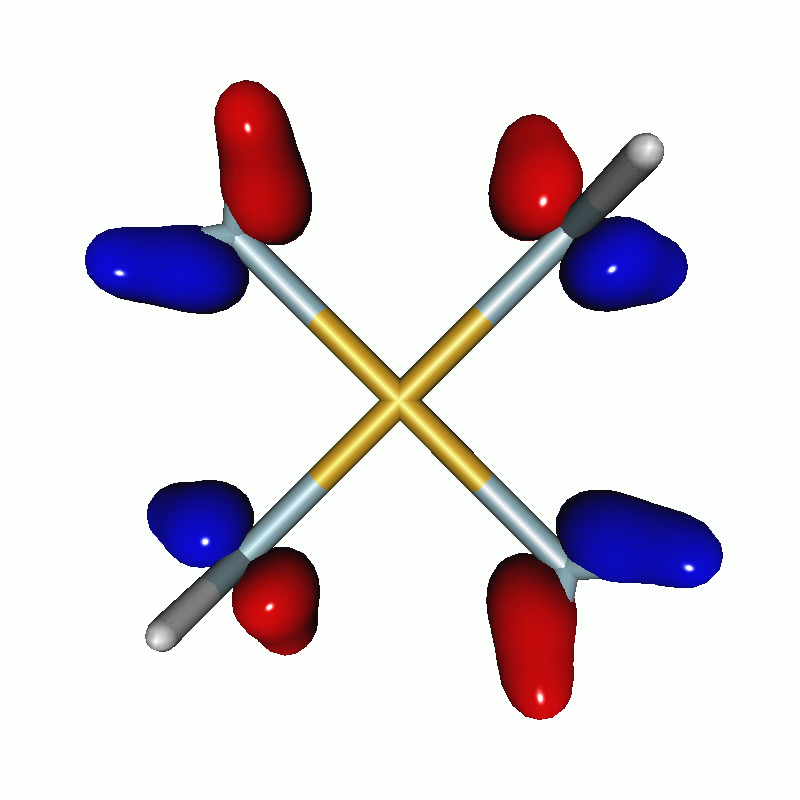}}
\caption{SOMO $S$ at the ROHF level (a) and ligand donor orbital $L$ (b) in the  $^2B_1$ excited state of the $\cunh$ molecule. }
\label{orb_cunh_e1}
\end{figure}

\begin{figure}[h]
\subfloat[SOMO]{\includegraphics[scale=0.25]{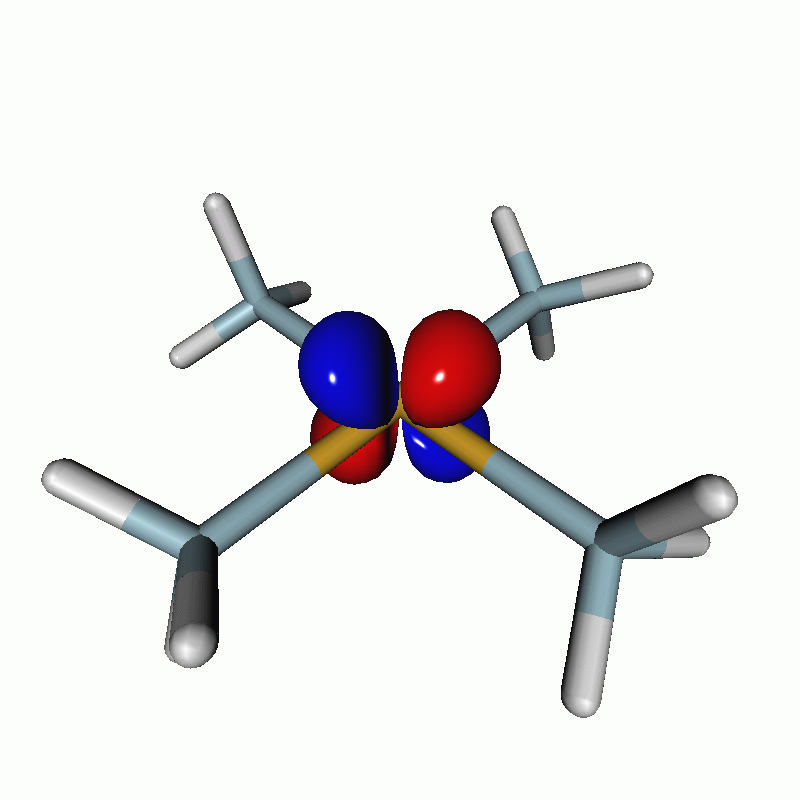}}
\subfloat[Donor orbital]{\includegraphics[scale=0.25]{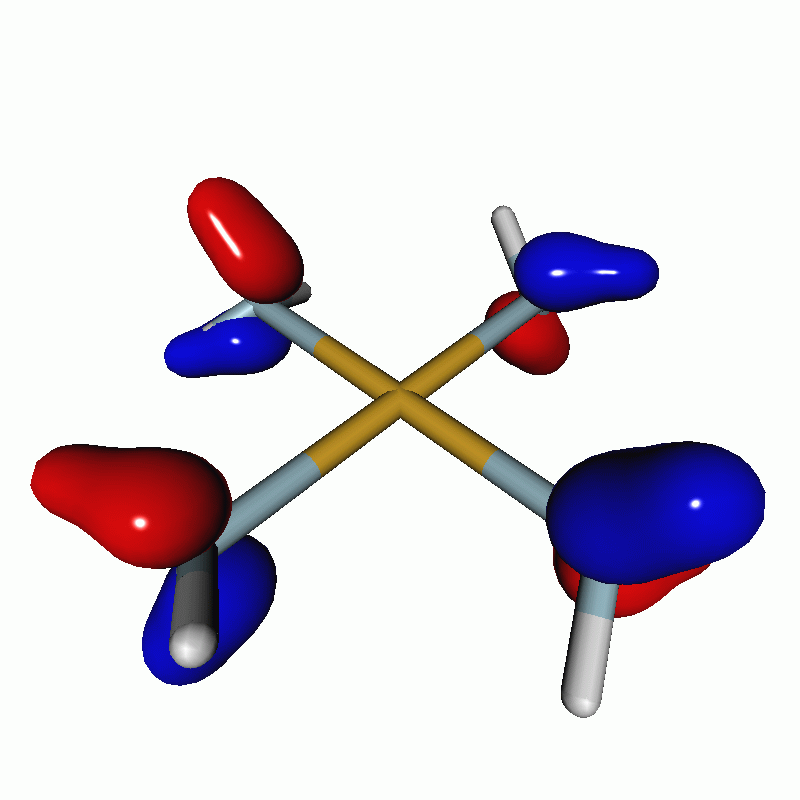}}
\caption{SOMO $S$ at the ROHF level (a) and ligand donor orbital $L$ (b) in the  $^2E$ excited state of the $\cunh$ molecule. }
\label{orb_cunh_e2}
\end{figure}

Having established the accuracy and reliability of FOBOCI, this greatly simplified wavefunction can be analysed in detail
to gain further insight into the relative levels of correlation-induced spin-delocalisation among the low-lying states.
We use the M\o ller-Plesset perturbation series for this purpose, which
corresponds to a Taylor expansion of the CI equations in this subspace.
At second order, neglecting exchange integrals, and the singles contribution,  the FOBOCI energy is
\begin{equation}
 \label{e2}
 e^{(2)} = \sum_{i,a} c_{Li}^{Sa\,(1)} \elemm{\rm ROHF}{H}{_L ^{S}\, _i ^a} \approx \sum_{i,a} \frac{(SL|ia)^2}{\epsilon_L - \epsilon_S+ \epsilon_i  - \epsilon_a}
\end{equation}
\begin{figure}[h]
\includegraphics[scale=0.75]{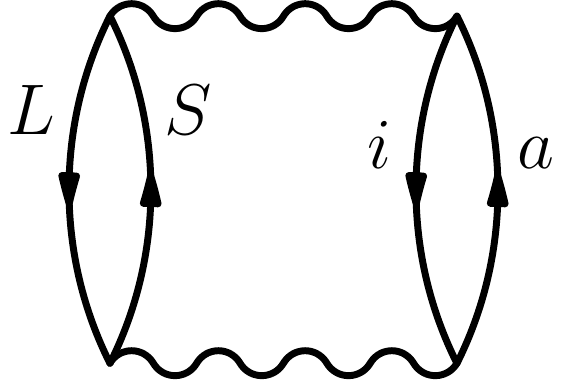}
\caption{Main diagrams involved in the calculation of $e^{(2)}$ within the FOBOCI space. }
\label{Fig_14_diagrams}
\end{figure}
The diagramatic representation is displayed in Figure \ref{Fig_14_diagrams}.
As previously noted, the second-order energy already shows a differential role between the ground state and the excited states. 
The electrostatic interaction between the SOMO $S$ and donor ligand $L$ orbitals dictates the crystal field splitting
and is therefore larger in the ground state than in the excited states. The integrals $(SL|ia)$ are correpondingly larger
in the ground state, resulting in a larger correlation energy, which raises the electronic transition energy.
Figures \ref{orb_cucl4}, \ref{orb_cuwater}, \ref{orb_cunh3}, \ref{orb_cucl4_e1}, \ref{orb_cucl4_e2}, \ref{orb_cuwater_e1}, \ref{orb_cuwater_e2}, \ref{orb_cunh_e1} and \ref{orb_cunh_e2} depict the SOMO and ligand donor orbitals.

The first contribution to the energy from the LMCT determinant is obtained at fourth order: the second order LMCT coefficient modifies the coefficients of the double excitations at third order. 
Neglecting minor exchange contributions, the second order coefficient $c_{\rm LMCT}^{(2)} $ is 
\begin{equation}
 c_{\rm LMCT}^{(2)} \approx \sum_{ia} \frac{(SL|ia)}{\epsilon_L - \epsilon_S+ \epsilon_i  - \epsilon_a} \frac{(ia|SS) - (ia|LL)}{\epsilon_L - \epsilon_S}
\end{equation}
\begin{figure}[h]
\subfloat[]{\includegraphics[scale=0.75]{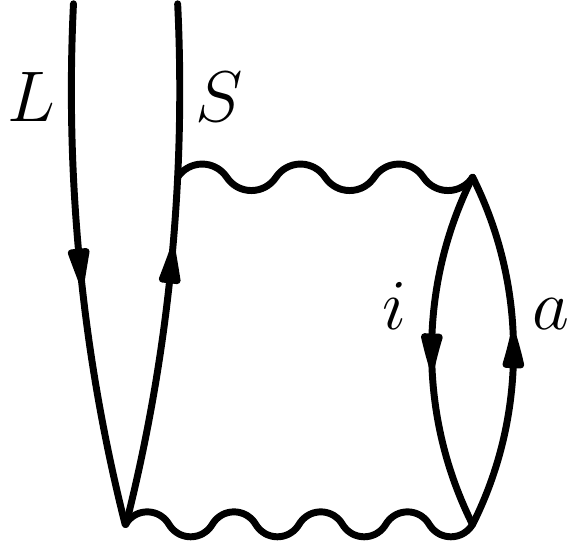}}
\subfloat[]{\includegraphics[scale=0.75]{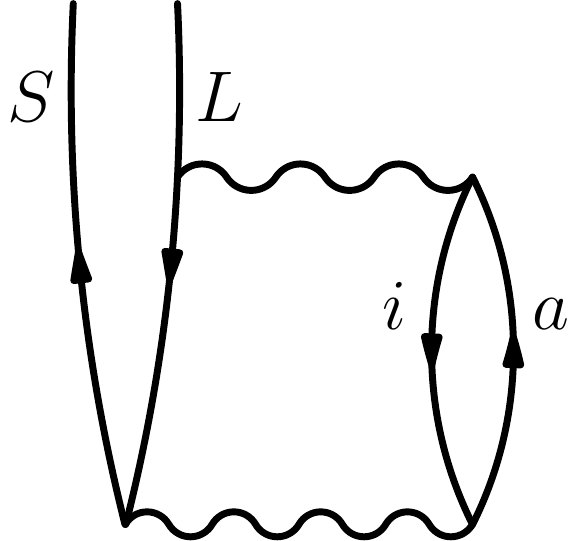}}
\caption{Main diagrams involved in the calculation of $c_{\rm LMCT}^{(2)} $ within the FOBOCI space. (a) corresponds to $\frac{(SL|ia)}{\epsilon_L - \epsilon_S+ \epsilon_i  - \epsilon_a} \frac{(ia|SS) }{\epsilon_L - \epsilon_S}$ and (b) to $ - \frac{(SL|ia)}{\epsilon_L - \epsilon_S+ \epsilon_i  - \epsilon_a} \frac{(ia|LL)}{\epsilon_L - \epsilon_S}$}
\label{c2_diagrams}
\end{figure}
The diagramatic representation is displayed in Figure \ref{c2_diagrams}.
Since the integrals $(SL|ia)$ are larger in the ground state than the excited state, $c_{\rm LMCT}^{(2)} $ is also larger for the ground state.
The full fourth-order energy expression is involved even for the FOBOCI space. The part that can be directly compared to the second-order energy is dominant and is given by 
\begin{equation}
 \label{e2_4}
 e^{(2)} + e^{(4)} \approx  \sum_{i,a} \frac{(SL|ia)^2}{\epsilon_L - \epsilon_S+ \epsilon_i  - \epsilon_a} \, \left( 1+ \frac{\left((ia|SS) - (ia|LL)\right)^2}{(\epsilon_L - \epsilon_S+ \epsilon_i  - \epsilon_a)(\epsilon_L - \epsilon_S)} \right)
\end{equation}
\begin{figure}[h]
\subfloat[]{\includegraphics[scale=0.75]{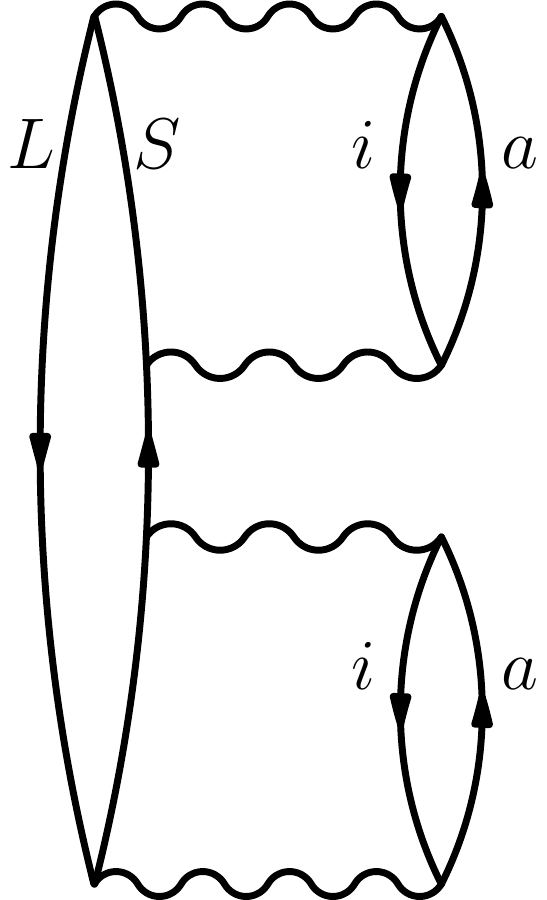}}\qquad
\subfloat[]{\includegraphics[scale=0.75]{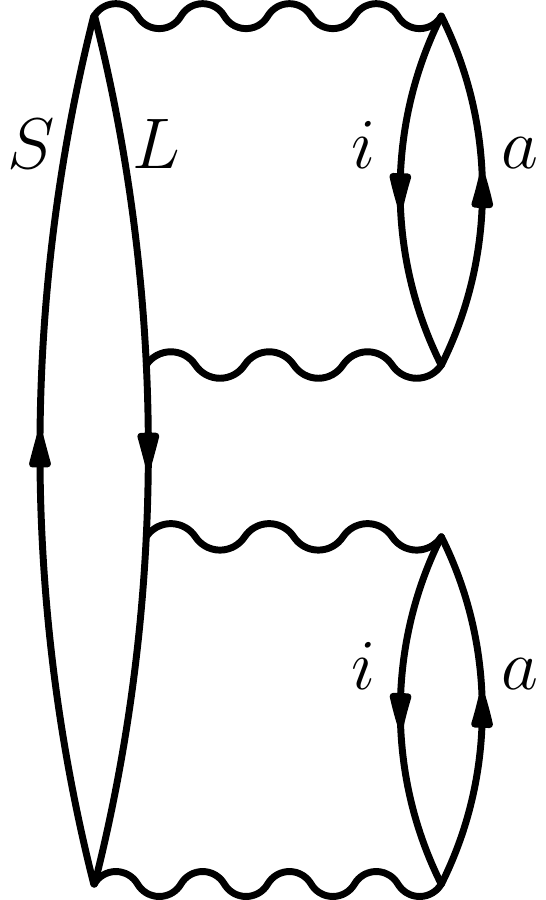}}\\
\subfloat[]{\includegraphics[scale=0.75]{Fig_18.pdf}}\quad
\subfloat[]{\includegraphics[scale=0.75]{Fig_17.pdf}}
\caption{Main diagrams involved in the calculation of $e^{(4)}$ within the FOBOCI space. 
(a) corresponds to 
$\frac{(SL|ia)^2}{\epsilon_L - \epsilon_S+ \epsilon_i  - \epsilon_a} \, \frac{(ia|SS)^2}{(\epsilon_L - \epsilon_S+ \epsilon_i  - \epsilon_a)(\epsilon_L - \epsilon_S)}$, 
(b) corresponds to 
$ \frac{(SL|ia)^2}{\epsilon_L - \epsilon_S+ \epsilon_i  - \epsilon_a} \, \frac{(ia|LL)^2}{(\epsilon_L - \epsilon_S+ \epsilon_i  - \epsilon_a)(\epsilon_L - \epsilon_S)}$, 
and (c) and (d) to 
$ - \frac{(SL|ia)^2}{\epsilon_L - \epsilon_S+ \epsilon_i  - \epsilon_a} \, \frac{(ia|SS)(ia|LL)}{(\epsilon_L - \epsilon_S+ \epsilon_i  - \epsilon_a)(\epsilon_L - \epsilon_S)}$. 
}
\label{e4_diagrams}
\end{figure}
The diagramatic representation of the approximation to $e^{(4)}$ is displayed in Figure \ref{e4_diagrams}.
As the energy denominators are always negative, and the numerators always positive, 
the higher-order effects enhance the second order energy correction through an interaction between LMCT and the double excitations $\ket{_L ^{S}\, _i ^a}$.
The differential effects at second order are magnified by the LMCT, which explains 
the importance of both the LMCT and the double excitations  $\ket{_L ^{S}\, _i ^a}$ in the correct prediction of the electronic transition energies.
This perturbation perspective can be connected to the VB picture through a decomposition in terms of strongly localised orbitals. This
analysis is somewhat involved, but the conclusions are that two physical effects at work: a small dispersive interaction between the
ligand lone pairs and the electron in the SOMO; and a comparatively large breathing orbital relaxation induced by the LMCT VB component.

\section{Single-reference methods}
\label{sec_single}

The success of the FOBOCI shows that very reasonable descriptions of the wavefunctions, electronic transitions and spin densities can be obtained 
at reduced computational cost through a careful selection of the CI space. The excitation manifold in FOBOCI is a subset of CISD, since it
contains all single excitations and a specific class of double excitations, and CISD may be anticipated to be even more accurate.
This section is dedicated to the investigation of the performance of single-reference wave function based methods.

\subsection{CISD and CCSD: the size extensivity error}

The results of the CISD calculations are reported in Tables \ref{results_cu}, \ref{table_cu} and \ref{results_cu_sd} for 
the electronic transitions, the amplitudes of the LMCT and the spin density, respectively. Contrary to expectation, 
CISD performs systematically worse than FOBOCI, underestimating the electronic transitions by at least 5 mH, with a 
corresponding underestimation of the amplitudes of the LMCT by at least a factor of two. 
The results of CCSD calculations are also reported. CCSD is in much better agreement with the near FCI results, with
a maximum error of only 1.6 mH for the transition energies. In the discussion below we demonstrate that the failure
of CISD is a direct consequence of the lack of size extensivity of the CISD wavefunction and energy.

The CISD and CCSD equations can be directly compared when using the \textit{unlinked} CCSD formalism.
In both CISD and CCSD, discarding the spin polarisation energy from the Brillouin terms, the correlation energy is
\begin{align}
 \label{ecorr}
 E_{\rm corr} &= \sum_{jkbc} \elemm{\rm ROHF}{H}{_{jk}^{bc}} \,\, c_{jk}^{bc} \\
              &= E^{L}_{\rm corr}(LiSa) + E^{UL}_{\rm corr}(LiSa)
\end{align}
In the second line we have introduced a decomposition into \textit{linked} and \textit{unlinked} contributions \textit{with respect to} 
to a particular double excitation $\ket{_{Li}^{Sa}}$.
The \textit{unlinked} correlation energy $E^{UL}_{\rm corr}(LiSa)$ is the sum over all quadruplet of indices $(j,k,b,c)$ in eq. \ref{ecorr} which \textit{do not} match any of the four indices $(L,i,S,a)$,
whereas the \textit{linked} part $E^{L}_{\rm corr}(LiSa)$ is the sum over all quadruplet of indices $ (j,k,b,c)$ in eq. \ref{ecorr} which match \textit{at least} one of the four indices $(L,i,S,a)$.
The CISD equation for the coefficient for the double excitations into $\ket{_{Li}^{Sa}}$ is
 \begin{align}
 c_{Li}^{Sa}  &= \frac{1}{\Delta_{Li}^{Sa}} \Big[ \elemm{_{Li}^{Sa}}{H}{\rm ROHF} + \sum_{jb} c_{j}^b \elemm{_{Li}^{Sa}}{H}{_j^b} + \sum_{jkbc} c_{jb}^{bc}  \elemm{_{Li}^{Sa}}{H}{_{jk}^{bc}}\Big] \\
 \Delta_{Li}^{Sa} &=  E_{\rm ROHF} - \elemm{_{Li}^{Sa}}{H}{_{Li}^{Sa}}  + E^{L}_{\rm corr}(LiSa) + E^{UL}_{\rm corr}(LiSa)
 \end{align}
whereas the corresponding CCSD equation is
 \begin{align}
 c_{Li}^{Sa}  &= \frac{1}{\Delta_{Li}^{Sa}} \Big[ \elemm{_{Li}^{Sa}}{H}{\rm ROHF} + \sum_{jb} c_{j}^b \elemm{_{Li}^{Sa}}{H}{_j^b} + \sum_{jkbc} c_{jb}^{bc}  \elemm{_{Li}^{Sa}}{H}{_{jk}^{bc}} \nonumber \\ 
& \qquad \qquad + \sum_{I \in T,Q} \{\elemm{_{Li}^{Sa}}{H}{I} \ovrlp{I}{\rm CCSD}\}_L \Big] \\
 \Delta_{Li}^{Sa} &=  E_{\rm ROHF} - \elemm{_{Li}^{Sa}}{H}{_{Li}^{Sa}}  + E^{L}_{\rm corr}(LiSa)
 \end{align}
where $T,Q$ are triple and quadruple excitations and the $c_{Li}^{Sa} E^{UL}_{\rm corr}(LiSa)$ term from
the denominator exactly cancels the
\textit{unlinked} parts of $\sum_{I \in T,Q} \elemm{_{Li}^{Sa}}{H}{I} \ovrlp{I}{\rm CCSD}$. 
In the CCSD equations, only the \textit{linked} correlation energy survives in the demoninator, 
whereas the total correlation energy remains in the denominator of the CISD equations. 
For the copper complexes, the total correlation energy is in the order of 10 eV, and the
\textit{unlinked} component accounts for more than $95\%$.
The presence of $E^{UL}_{\rm corr}(LiSa)$ in the CISD equations therefore introduces a spurious 10 eV shift 
in all energy denominators, which dramatically reduces the coefficients of the double excitations $c_{Li}^{Sa}$. 
Consequently, the correlation energy in general, and the differential correlation effects arising from
$c_{Li}^{Sa}$ in particular, are systematically underestimated at the CISD level, resulting in poor transition energies.

Tables \ref{results_cu}, \ref{table_cu} and \ref{results_cu_sd} also report the results from 
the CISD(SC)$^2$ method,\cite{meller-sc2,malrieu-sc2-2,Malrieu-SC2} where the CISD equations are modified by
removing $E^{UL}_{\rm corr}(LiSa)$ from the denominator.
CISD(SC)$^2$ repairs the errors of CISD and indeed performs comparably to CCSD, indicating that the
higher-order linked terms that are missing from CISD(SC)$^2$ do not play a large role in the energy differences between the excited states.
To conclude this analysis, we note that FOBOCI does not contain unlinked terms with respect to $\ket{_{Li}^{Sa}}$.
The FOBOCI correlation energy can be expressed as
\begin{equation}
 \label{ecorr_fobo}
 E_{\rm corr}^{\rm FOBOCI} = \sum_{jb} c_{Lj}^{Sb} \elemm{\rm ROHF}{H}{_{Lj}^{Sb}}
\end{equation}
and is therefore always {\it linked} with respect to any double excitation $\ket{_{Li}^{Sa}}$ present in the FOBOCI.
The FOBOCI also therefore does not suffer from size inconsistency errors for the terms that dominante
the differential correlation effects among the low-lying electronic states.

\subsection{CCSD(T) and BCCD(T)}
Table \ref{results_cu} reports the transition energies computed at the CCSD, BCCD, CCSD(T) and BCCD(T) levels using the 6-31G basis set.
The Brueckner coupled cluster results are included since we have shown that orbital relaxation effects are important and
the spin densities on the copper atom from the Brueckner orbitals are listed in Table \ref{results_cu_sd}.
In the 6-31G basis set, the CCSD, BCCD, CCSD(T) and BCCD(T) results are all close to the CIPSI values, with nothing to 
significantly favour one method over the other. The 6-31G basis is too small to reliably assess the importance of triple excitations,
so we performed additional CIPSI and coupled-cluster calculations where $f$ polarization functions are added to the copper atom, which
we denote the 6-31G*(Cu) basis. This was only feasible for the $\cucl$ molecule and the results are collected in Table \ref{results_cu_bis}.
The necessity for three-body correlation in approaching near FCI quality transition energies is clearly apparant.
Unfortunately, it is not possible to comment on the relative merits of BCCD(T) over CCSD(T) based on these numerical tests.

\begin{table}
\begin{center}
  \caption{Excitation energies (mH) at various levels of calculations for the $\cucl$ molecule in the 6-31G*(Cu) basis set. }
\scalebox{0.8}{
\begin{tabular}{|l|ccccc|c|c|}
\hline
Electronic transition          & ROHF  & CCSD      & BCCD    &  CCSD(T)  &BCCD(T)&  FOBOCI     & CIPSI     \\
\hline                                                                                                                          
$^2B_{1g}$  $-$ $^2B_{2g}$     & 30.2  & 45.7      & 45.1    &   46.7    & 46.7  &   44.0      & 46.8(1)    \\
$^2E_g$     $-$ $^2B_{2g}$     & 38.6  & 54.9      & 54.3    &   56.0    & 56.0  &   52.5      & 55.6(1)    \\
\hline                                                                                                                                                                       
\end{tabular}
\label{results_cu_bis}
}
\end{center}
\end{table}

\section{Basis set limit CCSD(T) and BCCD(T) transition energies}
\label{sec_references}
The success of CCSD(T) and BCCD(T) in reproducing near FCI transition energies in small basis sets, encourages us to 
use these methods to obtain high-quality reference values near the basis set limit.
Table \ref{ccf12_cu} reports the results of ROHF-UCCSD(T)$_\text{(F12*)}$ and UBCCD(T)$_\text{(F12*)}$ 
calculations\cite{hattig_ccf12,tew-os,tew_bcc} using the aug-cc-pwCVTZ-DK basis sets.
The X2C method\cite{X2C} was used to account for scalar relativistic effects and an exponent of 1.2 $a_0^{-1}$ was used in the F12 correlation factor.\cite{tew_f12} Table \ref{ccf12_cu} lists the additive contributions from the CCSD, (T), the CABS singles correction\cite{tew-Fig_14,tew_bcc} and the F12 correction
for frozen core calculations, where an argon core was used for copper, and a helium core for oxygen and nitrogen. The core-valence correlation correction
is the difference between the full valence only ROHF-UCCSD(T)$_\text{(F12*)}$ or UBCCD(T)$_\text{(F12*)}$ calculation and calculations correlating all 
electrons except the neon core at the copper. 

Concerning the dependence of the excitation energies on basis set, we find that while the values differ substantially from those computed with a 6-31G basis,
the CABS singles correction is negligable and the ROHF energies are converged to within 0.2 mH at the aug-cc-pwCVTZ-DK level. 
The F12 contribution is also less than a mH, suggesting that short-range dynamical correlation is not decisive in the ordering of the excited states
and that the CCSD(T) and BCCD(T) values are well converged with respect to one-particle basis set size at the aug-cc-pwCVTZ-DK level.
We note that since the F12 correction is based on the cusp condition for the
first-order amplitudes,\cite{bokhan-os,tew-os} it contains contributions of the type $f_{12}\vert Li \rangle$, 
but misses F12 contributions of the type $f_{12}\vert Sa \rangle$, and can therefore be expected to give slightly too low excitation energies 
with small basis sets. The magnitude of this effect, however, would appear to be very small, particularly in the Brueckner calculations where the 
orbital optimisation reduces this bias considerably. 
We ascribe a basis set incompleteness error bar of 0.5 mH for CCSD(T)$_\text{(F12*)}$ excitation energies, and 0.2 mH for BCCD(T)$_\text{(F12*)}$ 
excitation energies. 

Concerning the dependence of the excitation energies on the level of correlation treatment, we find that BCCD systematically predicts lower
transition energies than CCSD. This is pattern is reversed when comparing CCSD(T) and BCCD(T). 
Although the (T) energy is smaller for BCCD than for CCSD, the differential effect of the (T) triples correction on the excitation energies is larger
for BCCD than for CCSD. While the inclusion of high-order orbital relaxation effects in BCCD would favour this method, the (T) correction is anticipated to be biased to the ground state in both cases. Without benchmark calculations in a larger basis set, it is difficult to be sure which method is superior. 
We therefore quote the UBCCD(T)$_\text{(F12*)}$ as reference values for the transitions and assume that the difference between the ROHF-UCCSD(T)$_\text{(F12*)}$ and UBCCD(T)$_\text{(F12*)}$ values is a minimum estimate for the error bar.

The ab initio reference values are not expected to agree perfectly with experimental values, since the former are gas phase data and the latter are
obtained from electronic absorption spectroscopy of single crystals containing the gas phase chromophore, which are subject to crystal field effects and
geometric relaxation. For $\cucl$ Solomon and coworkers estimated the effect of crystal lattice on excitation energies from lattice model calculations at the level of DFT calculations,\cite{solomon_cucl4} reporting that it is at most 5 mH. The agreement between theory and experiment is within this error bar.
The experimental values for $\cunh$ in our table differ from those in other theoretical works\cite{neese_CuNH3_SOS,g_tensor_pierloot_mscaspt2}, where
the values for the electronic transitions to the ${}^2B_1$ and ${}^2E$ states are 63.8 and 79.7 mH, respectively.
With these values, however, the discrepancy between theory and experiment, however, is much larger than expected, as indeed previously noted by Neese\cite{neese_CuNH3_SOS}. The reason for this discrepancy is that the experimental values were measured by Hathaway and co-workers in 1969 for single crystals of Na$_4$Cu(NH$_3$)$_4$[Cu(S$_2$O$_3$)$_2$], which was assumed to have a square planar Cu(NH$_3$)$_4$$^{2+}$ environment. However, prompted by Morosin's more accurate X-ray data\cite{Morosin-cunh3_4}, Hathaway published a revised crystal structure interpretation indicating that the experiments were actually performed on a crystal with a weakly coordinating mono-ammonia adduct Na$_4$Cu(NH$_3$)$_4$[Cu(S$_2$O$_3$)$_2$],NH$_3$ that has a time average stereochemistry at the Cu atom of a tetragonal-octahedron.\cite{Hathaway-cunh3_4} In that same work, the electronic spectrum of Na$_4$Cu(NH$_3$)$_4$[Cu(S$_2$O$_3$)$_2$],H$_2$O was reported and analysed, and shown to have an effective square-planar CuN$_4$ stereochemistry with a freely rotating water molecule in the pocket at [0,0,$\tfrac12$]. The electronic transitions meausred were 83.8 and 87.5 mH to the ${}^2B_1$ and ${}^2E$ states, respetively. We therefore use these values in our table and indeed they agree with our computed values to within 5 mH, which can be attributed to be largely from small structural and environmental effects.

\begin{table}
\begin{center}
\caption{CC excitation energies (mH) for the $\cucl$, $\cunh$ and $\cuwater$ molecules. }
\scalebox{0.65}{
\begin{tabular}{|l|cccccccccccccc|}
\hline
Electronic transition   & ROHF  & CCSD  & $\Delta$HF & $\Delta$F12 & $\Delta$(T) & $\Delta$CV & CCSD(T)-F12 & BCCD & $\Delta$Ref & $\Delta$F12 & $\Delta$(T) & $\Delta$CV &  BCCD(T)-F12 & Exp.\footnotemark\footnotemark \\
\hline   
\multicolumn{15}{|l|}{$\cucl$}\\
\hline      
$^2B_{1g}$ $-$ $^2B_{2g}$   &  34.0  &  54.7  &  -0.1  &  -0.8   &  2.9  &  0.4  &  57.2  &   53.2  &   -0.1   &  -0.4  &  5.5  &  1.1  & 59.3  &  57.0 \\
$^2E_g$    $-$ $^2B_{2g}$   &  46.0  &  64.7  &  -0.1  &  -0.6   &  2.7  &  0.0  &  66.7  &   63.5  &   -0.2   &  -0.4  &  4.8  &  0.6  & 68.4  &  64.7 \\
\hline     
\multicolumn{15}{|l|}{$\cuwater$}\\
\hline      
$^2B_{1g}$ $-$ $^2B_{2g}$   & 44.4 & 56.0 &  0.0 &  0.2 &  1.8 & 0.7 & 58.2 & 55.1 &  0.0 & -0.1 & 2.8 & 0.9 & 58.7 & - \\
$^2E_g$    $-$ $^2B_{2g}$   & 47.8 & 55.3 &  0.1 &  0.0 &  0.9 & 0.8 & 57.1 & 54.9 &  0.0 &  0.0 & 1.4 & 0.9 & 57.3 & - \\
\hline 
\multicolumn{15}{|l|}{$\cunh$}\\
\hline                                                      
$^2B_1$    $-$ $^2B_2$     & 68.6  & 89.5  &  0.0 & -0.3  &  3.6  &  0.0  &  92.7  & 87.6 & -0.1 & 0.0 & 6.3 & 0.0 & 93.7 & 87.5 \\
$^2E$      $-$ $^2B_2$     & 53.8  & 76.9  &  0.0 & -0.3  &  3.6  &  0.0  &  80.2  & 75.0 & -0.1 & 0.0 & 6.3 & 0.0 & 81.2 & 83.8 \\
\hline
\end{tabular}
\footnotetext{Single crystal electronic absorption spectroscopy of square planar cupric chloride \cite{Neese-cucl4}} 
\footnotetext{Single crystal electronic absorption spectroscopy of Na$_4$Cu(NH$_3$)$_4$[Cu(S$_2$O$_3$)$_2$],H$_2$O  \cite{Hathaway-cunh3_4}}
\label{ccf12_cu}
}
\end{center}
\end{table}

\section{Conclusion}
\label{sec_conclu}

Through careful benchmarking and theoretical analysis, this work highlights that several key effects are at play in
in the correct theoretical determination of both the electronic spectroscopy and the ground state spin density of 
a series of square planar coordinated Cu$^{2+}$ complexes, namely the $\cucl$, $\cunh$ and $\cuwater$ molecules. 
Definitive reference energies and wavefunctions for the three low-lying spin states of each molecule, in a modest 6-31G basis, were obtained from 
near FCI calculations performed using the CIPSI selected CI method. Analysis of these states revealed
the prevalence of a specific excited configuration in all of the computed wavefunctions, which plays a 
vital role in the spin density and energies of the spin-states.
This configuration corresponds in all cases to a single excitation from the ROHF determinant 
where an electron is excited from a ligand-like orbital to the SOMO which is mainly localised on the 
central Cu$^{2+}$ ion.
A valence bond-like analysis shows that the these excitations can be identified as LMCT components of the ground state wave functions, 
which can therefore be thought of as a superposition of Cu$^{2+}$ and Cu$^+$ oxidation states.

A perturbation analysis of the coefficient of these Slater determinants in the ground and excited 
state wave functions revealed that these determinants arise predominantly
due to the so called breathing orbital effect, an orbital relaxation induced by the change in oxidation state at the Cu as
a consequence of correlating the electrons.
This effect plays a key role in the differential energies between the spin states and must be properly represented in the wavefunction
for a correct qualitative description of this class of systems.
Using these insights, we propose of a minimal CI space, the FOBOCI, which captures the key physical effects, and we
demonstrate numerically that it is able to reproduce quantitatively the energy differences and spin density of these three Cu$^{2+}$ complexes, 
even though it recovers less than 3$\%$ of the total correlation energy. 
The numerical evidence is further supported by a perturbational analysis, up to fourth-order in the energy, which, together with some
simple physico-chemical considerations, explains the success the FOBOCI in accurately describing the energy differences.

Having obtained detailed physical and mathematical insight into the theoretical description of these systems, we proceeded to
investigate the performance of the commonly applied wave function based methods, both single- and multi-reference.
Regarding multi-reference methods, the performance depends strongly on the choice of active space.
The minimal active space required for a qualitatively correct description is one containing the ligand orbital 
involved in the SOMO LMCT together with the double-$d$ shell, for the orbital relaxation.

Regarding single reference methods, we find that the correct description is obtained provided that
\begin{itemize}
 \item The wavefunction contains both the ROHF and SOMO LMCT configurations and all single excitations from each
 \item The wavefunction coefficients are obtained to at least 2nd order in perturbation theory (fourth order in the energy)
 \item The wavefunction coefficients are obtained in a size extensive manner
\end{itemize}
In this respect, our study reveals that CC-based methods are perfectly suited for the study of these Cu$^{2+}$ complexes,
since the excitation manifold of singles and doubles contains all important configurations, the iteratively 
optimised amplitudes correspond to high order in perturabtion theory, and the method is size extensive.
We find that CCSD(T) performs well despite exhibiting large $T_1$ and $D_1$ diagnostics for all wavefunctions. Indeed,
BCCD(T) and CCSD(T) return very similar results.
These diagnostics, based on the singles amplitudes, are large when there are strong orbital relaxation effects and are an 
indirect indication of multi-reference character at best. In this case the assumption that large $T_1$ and $D_1$ values
predict the failure of CCSD(T) is incorrect. 

Our study also reveals that CISD performs poorly. Our analysis proves that the non-size extensive nature of the
CISD equations leads to erroneous supression of correlating excitations, biasing spin states with smaller correlation energies.
We expect that our observation that size extensivity errors plague calculations of vertical spectrum of molecular complexes at equilibrium geometry
as well as dissociation energies will be generally applicable to all systems, since the errors simply grow with the magnitude of the
correlation energy.

Finally, having established the reliability of the CC-based methods for the determination of the energies of the spin states, 
we performed CCSD(T) and BCCD(T) calculations in a large basis set using explicitly correlated corrections in order to establish reference 
values for the energy differences of these three Cu$^{2+}$ complexes (see Table~\ref{ccf12_cu}). Our near basis set limit
core-valence correlated energies with scalar relativistic effects included agree with observed
energy differences from single crystal electronic absorption spectroscopy to within 5 mH, which is the same 
magnitude as the change expected due to placing the gas phase ion in the solid state crystal environment.

This study provides futher confirmation of the importance of LMCT in the determination of the properties of many $3d$ 
transition metal containing molecular complexes and highlights once more that metal-ligand delocalisation is very 
sensitive to the level to which electronic correlation is treated. 

\bibliography{paper}
\end{document}